\documentclass[prx,twocolumn]{revtex4}  

\usepackage{amssymb}
\usepackage{amsmath}
\usepackage{graphicx}
\usepackage{bbm}
\usepackage{color}
\usepackage{comment}
\usepackage{gensymb}
\usepackage{gensymb}
\usepackage{cancel} 
\usepackage{ulem}
\definecolor{blue}{rgb}{0,0,1}
\definecolor{grey}{rgb}{0.6,0.6,0.6}

\newcommand{\bra}[1]{\langle #1 |}
\newcommand{\ket}[1]{| #1 \rangle}

\newcommand{\pr}[1]{\textcolor{black}{#1}}

\begin{document}

\title{Toward high-fidelity quantum information processing and quantum simulation with spin qubits and phonons}

\author{I. Arrazola$^{1,2}$, Y. Minoguchi$^{2,3}$, M.-A. Lemonde$^4$, A. Sipahigil$^{5,6,7}$, and P. Rabl$^{2,8,9,10}$}

\affiliation{$^1$Instituto de F\'isica Te\'orica, UAM-CSIC, Universidad Aut\'onoma de Madrid, Cantoblanco, 28049 Madrid, Spain}
\affiliation{$^2$Vienna Center for Quantum Science and Technology, Atominstitut, TU Wien, 1040 Vienna, Austria}
\affiliation{$^3$Institute for Quantum Optics and Quantum Information -- IQOQI Vienna,
Austrian Academy of Sciences, Boltzmanngasse 3, 1090 Vienna, Austria}
\affiliation{$^4$Nord Quantique, Sherbrooke, Qu\'ebec, J1J 2E2, Canada}
\affiliation{$^5$Department of Electrical Engineering and Computer Sciences,
University of California, Berkeley, Berkeley, California 94720, USA}
\affiliation{$^6$Materials Sciences Division, Lawrence Berkeley National Laboratory, Berkeley, California 94720, USA}
\affiliation{$^7$Department of Physics, University of California, Berkeley, Berkeley, California 94720, USA}
\affiliation{$^8$Walther-Meißner-Institut, Bayerische Akademie der Wissenschaften, 85748 Garching, Germany}
\affiliation{$^9$Technische Universit\"at M\"unchen, TUM School of Natural Sciences, Physics Department, 85748 Garching, Germany} 
\affiliation{$^{10}$Munich Center for Quantum Science and Technology (MCQST), Schellingstr. 4, 80799 Munich, Germany} 

\date{\today}

\begin{abstract}
We analyze the implementation of high-fidelity, phonon-mediated gate operations and quantum simulation schemes for spin qubits associated with silicon vacancy centers in diamond. Specifically, we show how the application of continuous dynamical decoupling techniques can substantially boost the coherence of the qubit states while increasing at the same time the variety of effective spin models that can be implemented in this way.  Based on realistic models and detailed numerical simulations, we demonstrate that this decoupling technique can suppress gate errors by more than two orders of magnitude and enable gate infidelities below $\sim10^{-4}$ for experimentally relevant noise parameters. Therefore, when generalized to phononic lattices with arrays of implanted defect centers, this approach offers a realistic path toward moderate- and large-scale quantum devices with spins and phonons, at a level of control that is competitive with other leading quantum-technology platforms. 
\end{abstract}

\maketitle

\section{Introduction}

A significant amount of research effort is currently devoted to the development of quantum devices based on electronic and nuclear spin qubits associated with defect centers in solids. Combined with optical readout capabilities, the excellent coherence properties of the spins and their fixed location in a solid-state matrix make such systems an exquisite tool for various nano-scale sensing schemes~\cite{Rondin2014,Degen2017}, but also for coherent light-matter interfaces~\cite{Chu2014,Sipahigil2016,Trusheim2020,Merkel2020,Lukin2020} and building blocks for future quantum networks~\cite{Bernien2015,Evans2018,Nguyen2019,Ruf2021}. While most of the initial work in this direction was focused on the nitrogen-vacancy (NV) center in diamond~\cite{Doherty2013,Childress2014}, an equivalent level of control has now been demonstrated for many more defects in diamond and other materials~\cite{Wolfowitz2021}. However, in all those systems it is still an open challenge to identify reliable techniques for coupling two or more separated electron-spin qubits in a controlled  and scalable manner.  

Apart from direct magnetic~\cite{Dolde2013} and photon-mediated~\cite{Bernien2015,Sipahigil2016,Evans2018,Ruf2021} coupling schemes, where first important steps have already been experimentally demonstrated, the use of {\it phononic} quantum channels~\cite{Rabl2010,Habraken2012,Schuetz2015} is currently discussed as a promising alternative to overcome this scalability problem. While in a solid-state environment phonons are often a source of decoherence, the ability to fabricate high-Q mechanical resonators in diamond~\cite{Moller12,Ovartchaiyapong12,Burek13,Shandilya2022,Joe2023} and other materials offers completely new possibilities for using isolated phonon modes as a coherent quantum bus between spatially separated spin qubits. In the classical regime, it has already been demonstrated that static and dynamic strain fields in nanostructures provide versatile tools to control individual defect centers~\cite{MacQuarrie2013,Ovartchaiyapong2014,Golter2016,Meesala2016,Lee2017,Sohn2017,Whiteley2019,Chen2019,Maity2020,Wang2020}. It has further been proposed that at the level of single phonons, the same coupling mechanisms can be used to mediate spin-squeezing~\cite{Bennett2013,Li2020a}, to realize quantum gates~\cite{Xu2009,Albrecht2013,Qiao2020,Ren2022} and quantum communication protocols~\cite{Lemonde2018,Neuman2021}, or to engineer unconventional spin-spin and spin-phonon interactions~\cite{Dong2021,Li2020b,Dong2022,Shen2022} for quantum simulation applications. 

Among many possible defects, the negatively charged silicon-vacancy (SiV) center in diamond~\cite{Goss1996,Hepp2014,HeppThesis}  has been identified as a very promising candidate to implement such quantum spin-phonon interfaces~\cite{Lemonde2018,Meesala2018}. The electronic ground state of this center has both spin and orbital degrees of freedom, where the two lowest spin-orbit coupled states serve as a long-lived quantum memory~\cite{Sukachev2017}. At the same time, microwave fields or optical Raman beams can be used to drive transitions to higher orbital states~\cite{Lemonde2018}, which couple very strongly to phonons via strain. Hence, the level structure of this center provides a natural setting to obtain both strong and tunable spin-phonon interactions, which underlie the implementation of various quantum information processing and quantum simulation schemes. However, the theoretically estimated gate fidelities that are achievable with this basic coupling scheme are in the range of $\mathcal{F}\approx 0.9-0.99$, and potentially lower in actual experimental systems. While sufficient for many proof-of-concept demonstrations and basic entanglement operations, these values are not yet enough for high-fidelity and large-scale quantum information processing, for which significantly lower gate errors in the range of $ \mathcal{E} \approx 10^{-4}$ will be required.

In this work we propose and analyze improved protocols for phonon-mediated gate operation between SiV centers. The main purpose of this study is to show how the remaining limitations of this approach can be overcome and how gate fidelities in the range of $\mathcal{F}\simeq  0.999- 0.9999$ can be achieved under realistic conditions. To do so we consider, first of all, the direct coupling of the long-lived qubit states to phonon modes with frequencies of about 1-5 GHz~\cite{Meesala2018,Maity2020}. Compared to a Raman process involving phonon modes at $\sim50$ GHz~\cite{Lemonde2018}, this scheme relaxes the nanofabrication constraints for high-Q diamond resonators and it is also more convenient for accurate microwave control. However, for this direct interaction, the maximum achievable coupling strength is reduced, the spin-phonon coupling can no longer be tuned in situ and the temperature requirements to avoid thermal excitation of the phonon mode become more stringent. 

 To overcome these problems, we consider as a second ingredient the encoding of quantum information in dressed qubit states in the presence of a strong microwave field.  This driving field induces a built-in spin echo effect, which considerably extends the coherence time of the qubit states, as it has already been demonstrated for individual solid-state spins~\cite{Koppens2006,Xu2012,Golter2014,Barfuss2015,Laucht2017,Miao20} or other qubit systems~\cite{Ithier2005,Timoney2011,Yan2013}. Here we show that this decoupling technique is fully compatible with the phonon-mediated interactions between the SiV centers and can be further generalized to suppress thermal decoherence effects. It also enables a fast tunability of the effective qubit frequency, in order to control the coupling to the phonons with sufficiently high accuracy and to generate different types of spin-spin interactions for quantum simulation. We present detailed numerical calculations based on realistic system and noise parameters, and verify that these ingredients are sufficient to reach the desired operation accuracy. Therefore, although strong spin-phonon interactions at the quantum level have not been experimentally demonstrated yet, these findings lay out a realistic roadmap for a scalable and high-fidelity quantum processing platform with defect spin qubits in solids.

\section{A spin-phonon interface in diamond}\label{sec:Model} 
\begin{figure}
	\includegraphics[width=\columnwidth]{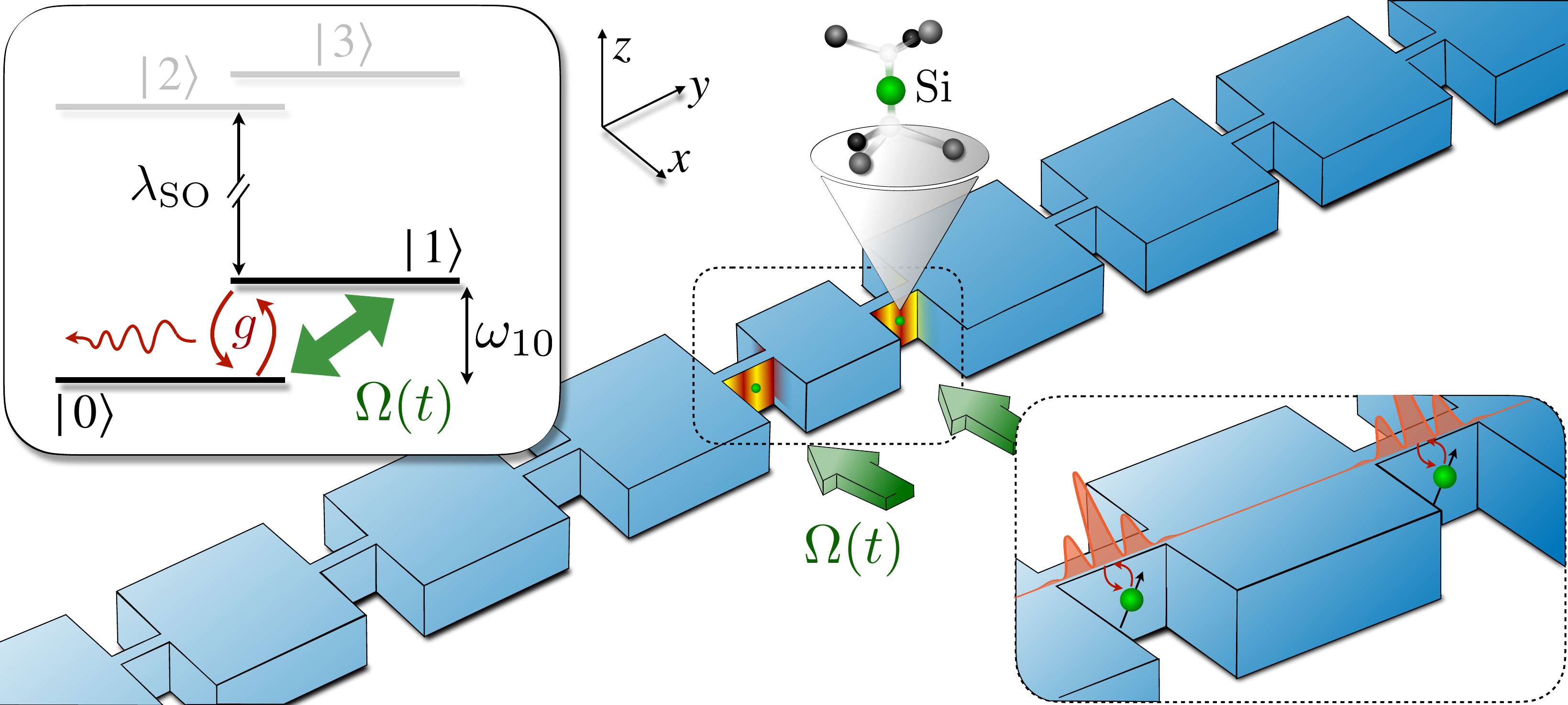} 
	\caption{Schematic of the setup. Two SiV centers are coupled to the strain of a quantized vibrational mode of frequency $\omega_{\rm ph}$, which is localized in a phononic crystal structure made out of diamond. The lowest states $|0\rangle$ and $|1\rangle$ in the ground state manifold of the SiV center are split by a frequency $\omega_{10}\sim \omega_{\rm ph}$, which results in a near-resonant spin-phonon interaction with coupling strength $g$. As indicated in the inset at the bottom, this coupling can be enhanced by identifying a phonon mode with a strain field that is concentrated in a small region around the SiV centers. In addition, a strong microwave field is used for a continuous dynamical decoupling of the spin qubits from low frequency noise, while preserving a strong phonon-mediated interaction (see text for more details).} 
	\label{Fig:Schema}
\end{figure}

We start our discussion by considering the basic setup depicted in  Fig.~\ref{Fig:Schema}, where $N=2$ SiV centers are coupled to a single isolated phonon mode of the surrounding diamond structure. In this setup, quantum states can be encoded in long-lived sublevels of the electronic ground state of the SiV centers~\cite{Sukachev2017}, while the phonon mode serves as a quantum bus to implement gate operations between the otherwise noninteracting defects. We model the whole system by the Hamiltonian  
\begin{equation} \label{eq:fullH}
H(t) = \sum_{i=1}^N H^{(i)}_{\rm SiV}(t)  + H_{\rm ph} + H_{\rm e-ph} + H_{\rm noise}(t),
\end{equation}
where $H_{\rm SiV}(t)$ describes the dynamics of a single SiV center under the influence of external control fields and $H_{\rm ph}=\hbar \omega_{\rm ph} a^\dag a$ is the Hamiltonian of the phonon mode with frequency $\omega_{\rm ph}$ and annihilation operator $a$. The phonon mode is coupled to both SiV centers via the strain interaction $H_{\rm e-ph}$. Finally, $H_{\rm noise}(t)$ accounts for the effect of low-frequency magnetic field fluctuations, which represent one of the dominating decoherence channels in this setup.

\subsection{The SiV ground state}
\pr{The SiV center is an anisotropic defect in diamond with a $D_{3d}$ point group symmetry and a high-symmetry axis aligned along the crystal axis $\langle 111 \rangle$}, which is assumed to lie along ${\bf e}_z$ in what follows. We are solely interested in the electronic ground state of this center, which is represented by an unpaired hole of spin $S = 1/2$ in two degenerate orbital states $|e_x\rangle$ and $|e_y\rangle$. In the presence of intrinsic strain and an external magnetic field, ${\bf B}_{\rm ext}(t)={\bf B}+ \mathcal{B}(t) \cos(\omega_d t){\bf e}_x$, the Hamiltonian for the ground state manifold reads
\begin{equation}
H_{\rm SiV}(t) =   H_{\rm SO} +   H_{\rm strain}  +   H_{\rm Z} + H_{\rm drive}(t). \label{Eq:HSiV}
\end{equation}
\pr{A more detailed derivation of this Hamiltonian can be found, for example, in Refs.~\cite{Goss1996, Hepp2014, HeppThesis}.} The first term in Eq.~\eqref{Eq:HSiV}, $H_{\rm SO} = -\hbar\lambda_{\rm SO}   L_z S_z$, is the spin-orbit coupling, where $L_z$ and $S_z$ are the $z$-components of the dimensionless orbital and spin angular momentum operators, ${\bf L}$ and ${\bf S}$. The spin-orbit coupling, $\lambda_{\rm SO}/(2\pi) \approx 46$ GHz, is the dominant energy scale and splits the four ground states into the two doublets $\{ \ket{e_-, \downarrow}, \ket{e_+, \uparrow}\}$ and $\{ \ket{e_+, \downarrow}, \ket{e_-, \uparrow}\}$. Here, $S_z|\!\!\uparrow,\downarrow\rangle=\pm1/2\ket{\!\!\uparrow,\downarrow}$, and $|e_\pm\rangle = (|e_x\rangle \mp i|e_y\rangle)/\sqrt{2}$ are the eigenstates of the orbital angular momentum operator,  i.e., $L_z|e_\pm\rangle =\pm |e_\pm\rangle$.

The bare spin-orbit doublets are further split and mixed by static strain in the host material, $H_{\rm strain}$,  and via the Zeeman effect, $H_{\rm Z}$. The strain only affects the orbital states and can be written as
\begin{equation} \label{Eq:Hstrain}
H_{\rm strain} = \hbar \left[(\Upsilon_x + i\Upsilon_y) L_- +  (\Upsilon_x - i\Upsilon_y)L_+\right], 
\end{equation}
where $L_+=L_-^\dag=\ket{e_+}\bra{e_-}$ induces transitions between the orbital states.
Here the coupling constants $\Upsilon_{x,y}/(2\pi)\sim 10$ GHz account for both the intrinsic Jahn-Teller effect as well as random lattice distortions, which will vary for every center. The Zeeman term, $ H_{\rm Z} =   g_e \mu_B {\bf B} \cdot   {\bf S} +   q \mu_B B_z L_z$, includes both the spin and the orbital Zeeman effect, where $g_e\simeq 2$ is the electron gyromagnetic ratio and $q\sim 0.1$ is the Ham reduction factor for the orbital states~\cite{Hepp2014, HeppThesis}. Finally, the last term in Eq.~\eqref{Eq:HSiV}, $H_{\rm drive}(t)= g_e \mu_B \mathcal{B}(t) \cos(\omega_d t-\phi_d) S_x$, accounts for the effect of an oscillating microwave field with frequency $\omega_d$ and phase $\phi_d$, which is used for the implementation of gate operations, as discussed below.  

\subsection{Qubit states}

In the absence of the driving field, $\mathcal{B}(t)=0$, the Hamiltonian $H_{\rm SiV}$ can be diagonalized, and in the following we identify the two lowest eigenstates with the qubit states $|0\rangle\approx \ket{e_-, \downarrow}$ and $|1\rangle\approx \ket{e_+, \uparrow}$ (see the inset in Fig.~\ref{Fig:Schema}), which are separated in frequency by $\omega_{10}$. In the examples below we consider a Zeeman splitting of about $\omega_{10}/(2\pi)\approx 3$ GHz, such that the qubits states are in resonance with a phonon mode of similar frequency. Under this condition, the remaining two states $|2\rangle\approx \ket{e_+, \downarrow}$ and $|3\rangle\approx \ket{e_-, \uparrow}$ are still far separated in energy and can be neglected in the dynamics.

\begin{figure}
	\includegraphics[width=\columnwidth]{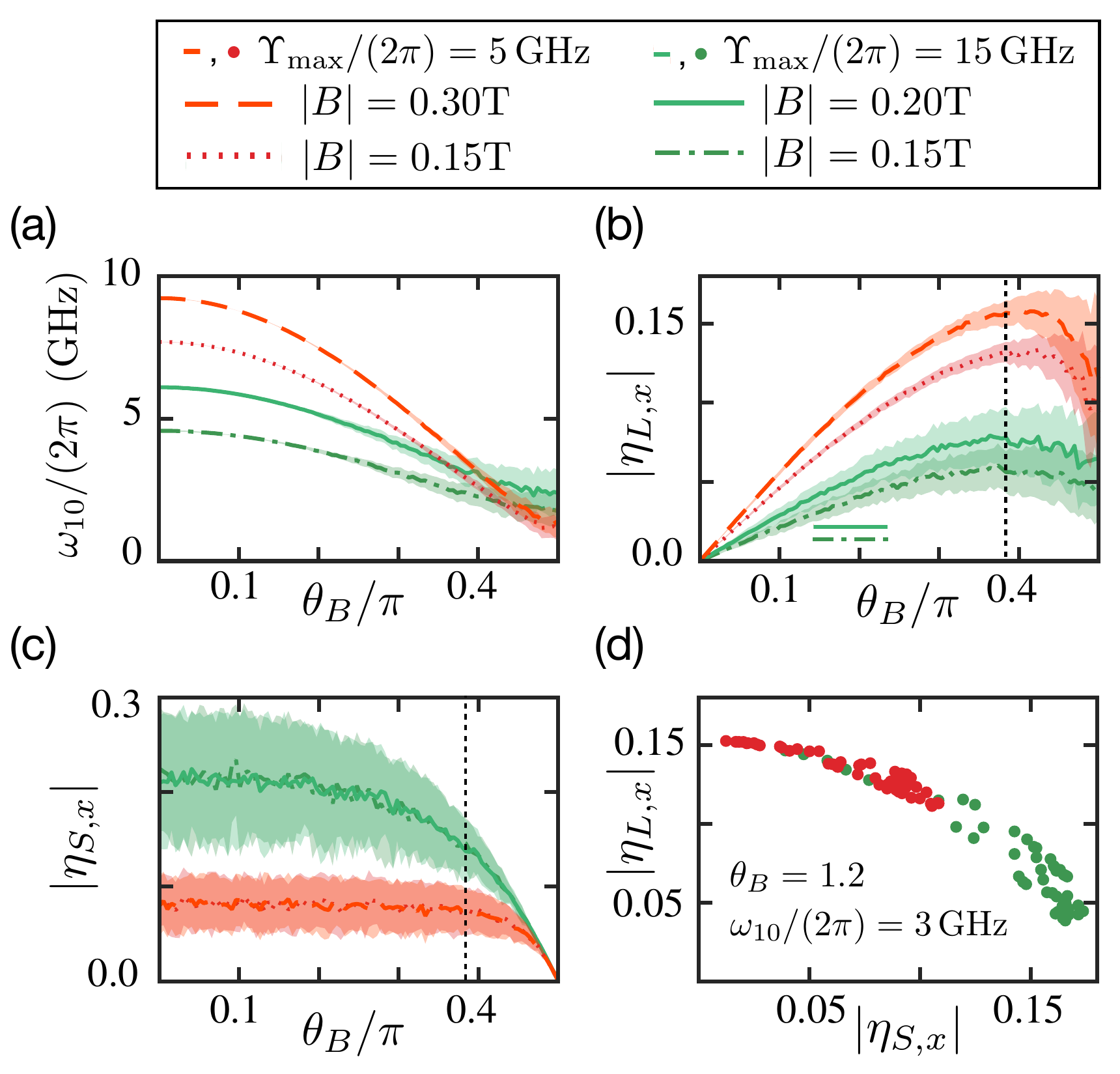} 
	\caption{Qubit parameters. Plots of (a) the qubit frequency splitting $\omega_{10}$, (b) the absolute values of the strain coupling matrix element $\eta_{L,x}$ and (c) the spin transition matrix element $\eta_{S,x}$ as a function of the rotation angle $\theta_B = \arctan(B_x/B_z)$ and different values of the static magnetic field, $|{\bf B}|$. In all plots, the respective quantities represent the average values obtained for a random distribution of static strain components $\Upsilon_{x,y}\in [-\Upsilon_{\rm max},\Upsilon_{\rm max}]$, and the shaded areas indicate their variation (one standard deviation). (d) For $\theta_B=1.2$ and randomly chosen $\Upsilon_{x,y}$ the value of $|{\bf B}|$ is adjusted to fix the qubit frequencies to $\omega_{10}/(2\pi)=3$ GHz. The plot shows the resulting distribution of matrix elements $\eta_{L,x}$ and $\eta_{S,x}$. }
	\label{fig2}
\end{figure}

Due to the orthogonality of both the orbital and the spin components, the bare spin-orbit states, $\ket{e_-, \downarrow}$ and $\ket{e_+, \uparrow}$, are neither coupled by phonons nor by the microwave driving field.  It is thus important to bear in mind that the actual qubit states are superpositions of all four basis states, with the mixing and phase angles depending on the static strain, $\Upsilon_{x,y}$, and the applied magnetic field, ${\bf B}$ (see Appendix~\ref{app:Eigenstates} for more details). Of specific relevance for the following discussion are the matrix elements
\begin{equation}
\eta_{S,x}= \langle 0|S_{x}|1\rangle, \qquad \eta_{S,z} =\langle 1|S_x|1\rangle - \langle 0|S_x|0\rangle,
\end{equation}
which describe the coupling of the qubit states to the magnetic driving field along ${\bf e}_x$, as well as the matrix elements
\begin{equation}
\eta_{L,x} = \langle 0|L_x|1\rangle, \qquad \eta_{L,z} =\langle 1|L_x|1\rangle - \langle 0|L_x|0\rangle,
\end{equation}
where $L_x=L_++L_-$. These quantities enter in the coupling to the quantized phonon mode, as discussed in Sec.~\ref{subsec:straincoupling}. In Fig.~\ref{fig2} we plot $|\eta_{S,x}|$ and $|\eta_{L,x}|$ together with the qubit frequency $\omega_{10}$ as a function of the static magnetic field angle $\theta_B = \arctan(B_x/B_z)$, assuming $B_y=0$. For these plots, the values for the static strain couplings $\Upsilon_{x,y}$ are randomly drawn from an interval $[-\Upsilon_{\rm max},\Upsilon_{\rm max}]$ and the solid lines (shaded areas) indicate the average values (variances) of the considered quantities. We see that for typical strain fields~\cite{Kehayias2019,Marshall2022}
and magnetic fields in the range of $|{\bf B}| \sim0.1-0.3$ T, we can tune the qubit frequency between $\omega_{10}/(2\pi) \approx 2-5$ GHz while achieving sizeable matrix elements of about $|\eta_{S,x}|\sim  |\eta_{L,x}|\sim 0.1$. This requires a static magnetic field applied mainly along the $x$-direction and either intrinsic or externally applied strain~\cite{Meesala2018}. By assuming that the qubits can be locally tuned to the same frequency of  $\omega_{10}/(2\pi) =3$ GHz, Fig.~\ref{fig2} (d) shows the expected distribution of matrix elements when we assume only a random intrinsic strain. The behavior of the less relevant matrix elements $\eta_{S,z}$ and $\eta_{L,z}$ is shown in Appendix~\ref{app:Eigenstates}.

Restricted to the two qubit states  $|0\rangle$ and $|1\rangle$ the Hamiltonian for a single SiV center reduces to ($\hbar=1$)
\begin{equation} \label{Eq:HSiVTLS}
\begin{split}
H_{\rm SiV}(t)= &H_q+\frac{\delta\omega(t)}{2}\sigma_z \cos(\omega_d t-\phi_d) \\
+  &\Omega(t)\left(\sigma_-e^{i\phi_S}+{\rm H.c.}\right)\cos(\omega_d t-\phi_d),
\end{split}
\end{equation}
where $H_q=\omega_{10}\sigma_z/2$, and we have introduced the Pauli operators $\sigma_- \equiv \ket0\bra1$ and $\sigma_z \equiv \ket1\bra1 - \ket0\bra0$. In Eq.~\eqref{Eq:HSiVTLS},
$\Omega(t) = |\eta_{S,x}| g_e \mu_B  \mathcal{B}(t)/\hbar$ is a time-dependent Rabi frequency, and $\delta\omega(t) = \eta_{S,z} g_e \mu_B  \mathcal{B}(t)/\hbar$. Note that here the complex phase factor in $\eta_{S,x}$ translates into $e^{i\phi_S}$, as $\eta_{S,x}=|\eta_{S,x}|e^{i\phi_S}$. In the following we are interested in conditions where $|\Omega(t)|,|\delta \omega(t)| \ll \omega_{d}\approx \omega_{10}$. In this case we can make a rotating wave approximation (RWA) and obtain the qubit Hamiltonian
\begin{equation} \label{Eq:drivenSiV}
H_{\rm SiV}(t)\simeq \frac{\omega_{10}}{2}\sigma_z + \frac{ \Omega(t)}{2}\left(\sigma_+e^{-i\omega_d t} + \sigma_-e^{i\omega_d t}\right),
\end{equation}
where we choose $\phi_d=\phi_S$ for convenience.
Note that by going from Eq.~\eqref{Eq:HSiVTLS} to Eq.~\eqref{Eq:drivenSiV} we have neglected off-resonant contributions, which can induce relevant corrections to the qubit dynamics. However, these corrections are deterministic, and in Appendix~\ref{app:LongitudinalCouplings} we show how they can be systematically taken into account without compromising the following predictions of gate fidelities obtained from Eq.~\eqref{Eq:drivenSiV}.

\subsection{Strain coupling to phonons}\label{subsec:straincoupling}
Apart from any static strain fields already included in $H_{\rm strain}$, the orbital states of the SiV center are also affected by the oscillating strain associated with the quantized phonon modes. The resulting electron-phonon interaction term adopts the same form as in Eq.~\eqref{Eq:Hstrain}, but with the corresponding strain fields $(\Upsilon_{x}, \Upsilon_{y}) \rightarrow (\upsilon_x, \upsilon_y)$ being quantized. More precisely, $\upsilon_x = d(\epsilon_{xx}-\epsilon_{yy}) + f\epsilon_{zx} $ and $\upsilon_y = -2d\epsilon_{xy} + f\epsilon_{yz}$ are proportional to the strain tensor elements defined as
\begin{equation}
\epsilon_{ij} = \frac{1}{2}\left( \frac{\partial u_i}{\partial x_j} + \frac{\partial u_j}{\partial x_i}\right). 
\end{equation}
Here $x_1 = x$ ($x_2 = y, x_3 = z$) and $\vec u$ denotes the quantized displacement field at the position of the SiV center. The strain susceptibilities for the SiV center, $d \approx -f \approx 1.5$ PHz, have been measured in Ref.~\cite{Meesala2018}.

In general, the quantized displacement field can be written as 
\begin{equation} \label{Eq:Disp}
\vec{u}(t,\vec{r}) = \sum_n \sqrt{\frac{\hbar}{2\rho \omega_n}} \vec{\zeta}_n(\vec{r})\left(a_ne^{-i\omega_n t}+{\rm H.c.}\right),
\end{equation}
where $a_n$ is the annihilation operator for a phonon with frequency $\omega_n$ and mode function $\vec{\zeta}_n(\vec{r})$. The latter is normalized to $\int d^3 r \,  \vec{\zeta}^*_n(\vec{r}) \cdot  \vec{\zeta}_m(\vec{r})=\delta_{n,m}$. For diamond nano-structures with dimensions  $\lesssim 5\,\mu$m, the spacing between the mode frequencies is $\Delta \omega \gtrsim 1$ GHz, which allows us to address only a single mode with frequency $\omega_{\rm ph}\approx \omega_{10}$ and mode function $\vec{\zeta}_{\rm ph}(\vec{r})$. Restricted to this mode and projected onto the qubit basis states $|0\rangle$ and $|1\rangle$, the spin-phonon coupling is given by 
\begin{equation}\label{Eq:JCHamiltonian}
H_{\rm e-ph} \simeq  \hbar \left(g a^\dag \sigma_- + g^* a \sigma_+\right),
\end{equation}
where we have already made the RWA and omitted the longitudinal coupling terms $\sim \eta_{L,z}$. Note that those contributions are not strictly negligible in the context of high-fidelity two-qubit interactions, but they can be reabsorbed into a redefinition of effective parameters and do not affect any of the main conclusions discussed below. Therefore, for the sake of clarity, we proceed with Eq.~\eqref{Eq:JCHamiltonian} and refer to Appendix~\ref{app:LongitudinalCouplings} for further details. The coupling constant $g$ depends on the details of the chosen phonon mode function, but can be written as
\begin{equation}
g=  d \epsilon_{\rm ph} \eta_{L,x},
\end{equation}
where $\epsilon_{\rm ph}$ is the characteristic strain amplitude per phonon. For a basic estimate, we consider the fundamental compression mode in a rectangular box of length $\ell$ and cross section $A$, with a frequency $\omega_{\rm ph}=c\pi/\ell$ and a maximal strain per phonon of
\begin{equation}
\epsilon_{\rm ph} \approx \sqrt{\frac{\hbar \omega_{\rm ph}}{ c^2  \rho \ell A}} .
\end{equation} 
Here $\rho\approx 3500$ kg/m$^3$ is the density and $c\approx 1.7\times 10^4$ m/s the speed of sound in diamond. For $\ell=3\, \mu$m,  $A=(200{\rm nm})^2$ and $|\eta_{L,x}|=0.1$, we obtain $\omega_{\rm ph}/(2\pi)\approx 3$ GHz,  $\epsilon_{\rm ph}\approx 4\times 10^{-9}$ and $g/(2\pi)\approx 0.6$ MHz. This coupling can be further enhanced by designing phononic crystals as depicted in Fig.~\ref{Fig:Schema}, where the strain field is strongly localized in an effective volume $V_{\rm eff}\ll \ell A$, while retaining approximately the same frequencies. Exact numerical simulations~\cite{Kuzyk18,Kim23} of such modes predict single-phonon strain amplitudes up to $\epsilon_{\rm ph}\approx 10^{-7}$ and, correspondingly, effective spin-phonon couplings up to $g/(2\pi)\approx 15$ MHz.

\subsection{Qubit-phonon interface}
In summary, by changing into a frame rotating with the driving frequency $\omega_d$, the coupling between the SiV centers and a single isolated phonon mode can be accurately described by a Jaynes-Cummings model of the form
\begin{equation}\label{eq:HQubitPhonon}
\begin{split}
	H(t) \simeq\, &  \hbar \sum_{i=1}^N \left[ \frac{\Delta}{2}\sigma_i^z +\frac{\Omega_i(t)}{2}\sigma_i^x \right]+  \hbar \Delta_{\rm ph} a^\dag a \\
	& + \hbar \sum_{i=1}^N \left(g_{i} a^\dag \sigma_i^- + g^*_{i} a \sigma_i^+\right) + H_{\rm noise}(t),
\end{split}
\end{equation}
where we have introduced the detunings $\Delta=\omega_{10}-\omega_d$ and $\Delta_{\rm ph}=\omega_{\rm ph}-\omega_d$.

Note that in the following we assume full local control of static and microwave fields in order to compensate for strain-induced inhomogeneities in the qubit frequencies and transition matrix elements. For a separation of the SiV centers of about $\sim 1\,\mu$m, this can be achieved by using submicron-scale gate electrodes, similar to what is used for the control of spin qubits in gate-defined quantum dots~\cite{Burkard23}.

Apart from the coherent dynamics and the magnetic field noise, we must also take into account the coupling of the phonon mode to its thermal surrounding. We do so by modeling the dynamics of the system density operator $\rho$ in terms of the master equation 
\begin{equation}\label{Eq:Master_Equation}
\dot{\rho}=-\frac{i}{\hbar}[H(t),\rho] +\kappa({n}_{\rm th}+1) \mathcal{D}[a]\rho +\kappa {n}_{\rm th} \mathcal{D}[a^\dagger]\rho.
\end{equation}
Here, $\mathcal{D}[a]\rho =a\rho a^\dagger -\frac{1}{2}\{a^\dagger a,\rho\}_+$, and $\kappa=\omega_{\rm ph}/Q$  and ${n}_{\rm th}=1/(e^{\hbar\omega_{\rm ph}/(k_{\rm B}T)}-1)$ are the decay rate and the thermal occupation number of the phonon mode for a given quality factor $Q$ and support temperature $T$. Note that phononic crystal resonators with mechanical 
quality factors $Q\gtrsim10^5$ have already been demonstrated~\cite{Joe2023} and in view of the continued advances in diamond nanofabrication, further improvements are expected.

\section{Dynamically protected spin-spin interactions} \label{sec:DressedQubits}
The Hamiltonian in Eq.~\eqref{eq:HQubitPhonon} is familiar from cavity QED systems with atoms and various other quantum technology platforms, such as trapped ions or superconducting circuits. In such settings, by choosing a large detuning, $|\Delta-\Delta_{\rm ph}|\gg g_i$ and $\Omega_i(t)=0$, the bosonic mode is only virtually populated, but mediates  an effective flip-flop interaction between the qubits. This interaction is enough to entangle the otherwise spatially separated spins or, more generally, implement a universal two-qubit gate. In this detuned regime and after optimizing the detuning,  the minimal gate error (see discussion below),
\begin{equation}\label{eq:minErrorUndriven}
\mathcal{E}_{\rm min}\approx  \frac{3}{2}\left(\frac{\pi}{4 \sqrt{\mathcal{C}}}\right)^{\frac{4}{3}} \sim \mathcal{C}^{-\frac{2}{3}},
\end{equation}
depends only on a single parameter, the so-called cooperativity 
\begin{equation}
\mathcal{C}= \frac{g^2 T_2^*}{\kappa(2n_{\rm th}+1)},
\end{equation}
where $T_2^*$ is the bare decoherence time of the spin qubits.  While for mechanical quality factors $Q=10^4-10^6$ and the couplings estimated above already very high values of $\mathcal{C}\sim10^3-10^4$ are theoretically achievable, it is not yet enough to reach gate errors below $10^{-3}$. One of the remaining sources of error comes from the short spin decoherence time, which in solids is typically limited to $T_2^*\sim 1-10\,\mu$s~\cite{Sukachev2017}, due to ubiquitous sources of magnetic field noise or other processes that cause slow but unknown frequency drifts.

To overcome dephasing in single-qubit control, one usually applies spin-echo and related dynamical decoupling sequences, where the evolution of the spin is interrupted by a discrete set of fast $\pi$-pulses~\cite{Sukachev2017}. These spin flips cancel any static frequency shifts, but also interfere with the coupling to the phonon mode and must be fast compared to all the relevant system timescales.  
Therefore, here we consider an alternative decoupling strategy, named continuous dynamical decoupling (CDD), where the spin is rotated all the time by applying a strong continuous driving field~\cite{Koppens2006,Xu2012,Golter2014,Barfuss2015,Laucht2017,Miao20,Ithier2005,Timoney2011,Yan2013}. Note that the use of CDD schemes has already been suggested in initial proposals for achieving strong spin-phonon interactions~\cite{Rabl09} and has also been analyzed in connection with cavity-mediated gate operations in different systems~\cite{Bermudez12,Cohen15,Cao17,Srinivasa23}, including experimental demonstrations with trapped ions~\cite{Tan13,Harty16} and superconducing circuits~\cite{Guo18}. Compared to those systems, the influence of quasi-static noise on defect centers in solids is even more detrimental, therefore, even greater benefits can be expected from CDD schemes. The purpose of the following analysis is to demonstrate that this is indeed the case and that this approach is sufficient to reach the targeted gate errors of $\lesssim 10^{-4}$ in a spin-phonon system.

\subsection{Low-frequency noise}\label{sub:lowfreq}
While the precise origin and details of the noise are often not known and may depend on the specific experimental setting, it has been demonstrated that  the bare dephasing time $T_2^*$ of SiV centers at low temperatures can be dramatically enhanced to values of about $T_2\approx 1$ ms by applying pulsed spin-echo techniques~\cite{Sukachev2017}. This observation shows that the decoherence of SiV qubits is primarily induced by slowly varying fluctuations of the transition frequency $\omega_{10}$, and for the following analysis it is enough to consider the noise Hamiltonian 
\begin{equation}\label{eq:dephasing}
H_{\rm noise} = \hbar \sum_{i=1}^N \frac{\xi_i(t)}{2}\sigma_i^z.
\end{equation}
Here the $\xi_i(t)$ are independent stochastic processes, which we model by a set of Ornstein-Uhlenbeck processes~\cite{Gillespie96} with zero mean and variance
\begin{equation}
\langle \xi_i(t)\xi_j (0)\rangle= \delta_{ij} \sigma^2e^{-t/\tau_c}.
\end{equation}
The parameters $\sigma$ and $\tau_c$ describe the strength of the noise and its correlation time, respectively. Within this noise model, 
we obtain 
\begin{equation}
T_2^*\simeq \frac{\sqrt{2}}{\sigma},   \qquad {\rm and}\qquad 
T_2\simeq  \sqrt[3]{\frac{12\tau_c}{\sigma^2}},
\end{equation}
which can be used to relate $\sigma$ and $\tau_c$ to the experimentally measurable values of the bare dephasing time  $T_2^*$ and the spin-echo coherence time $T_2$. From the parameters cited above we obtain typical values of $\sigma/(2\pi)\sim 100$ kHz and $\tau_c>1 $~s, indicating that the noise is to a good approximation quasi-static over the timescale of the gate operations considered in this work.

\subsection{Dressed qubit states}\label{sub:dressedqubits}
We now assume that the qubits are driven by a continuous microwave field, $\Omega_i(t)=\Omega$. In this case, the Hamiltonian of an individual SiV center can be written as  
\begin{equation}\label{eq:Single_Spin_Dressed}
H_q = \frac{\hbar \tilde \Omega}{2}\tilde{\sigma}_z + \frac{\hbar \xi(t)}{2} \left[\cos(\theta) \tilde{\sigma}_z + \sin(\theta) \tilde{\sigma}_x\right],
\end{equation}
where $ \tilde \Omega= \sqrt{\Delta^2 +\Omega^2}$ and the $\tilde \sigma_k$ are Pauli operators in the dressed qubit basis defined as 
\begin{eqnarray}
|\tilde 0\rangle &=&  \cos(\theta/2)|0 \rangle -  \sin(\theta/2) |1\rangle ,\\
|\tilde 1 \rangle &=&  \cos(\theta/2)|1\rangle +    \sin(\theta/2) |0\rangle,
\end{eqnarray}  
with a mixing angle $\theta$ given by $\tan(\theta)=\Omega/\Delta$. 

Equation~\eqref{eq:Single_Spin_Dressed} shows that increasing the ratio $\Omega/\Delta$ reduces the parallel noise component affecting the decoherence of the dressed qubit states, which should lead to an increase of the coherence time by
$T_2^*\rightarrow T_2^*/\cos^2(\theta)$. Obviously, this scaling does not hold near resonance, $\Delta=0$. In this limit, the noise is purely transverse, but because $\xi(t)$ is slowly varying, it cannot efficiently induce transitions between $|\tilde 0\rangle$ and $|\tilde 1\rangle$ when the splitting $\Omega\gg\sigma, \tau_c^{-1}$ is sufficiently large. The dominant effect of the noise then arises from a second-order shift
\begin{equation}\label{eq:SecondOrderNoise}
H_q \simeq \frac{\hbar \Omega}{2}\tilde{\sigma}_z + \frac{\hbar \xi^2(t)}{4\Omega} \tilde{\sigma}_z.
\end{equation}
This suppression of the noise by a factor of $\sigma/\Omega$ can be interpreted as a continuous spin-echo effect, where the original qubit states are continuously interchanged. 
 
Starting from an initial superposition state $|\psi_0\rangle =(|\tilde 0\rangle + |\tilde 1\rangle)/\sqrt{2}$, the subsequent evolution of the expectation value $\langle \tilde \sigma_- \rangle (t)$, averaged over all noise realizations, determines the loss of coherence of the dressed states. For general $\Delta$ and $\Omega$, the full form of this decoherence function has been derived in Ref.~\cite{Rabenstein04}, but in the limit of interest, $\Delta=0$ and $\tau_c\rightarrow\infty$, it simplifies to
\begin{equation}
\overline{\langle \tilde \sigma_- \rangle (t)} \simeq \sqrt{\frac{1}{1+i \sigma^2 t/\Omega}} \,\langle \tilde \sigma_- \rangle (0).
\end{equation}
From this expression we can identify a characteristic decoherence time
\begin{equation}
T_2^\Omega=  \frac{2 \Omega}{\sigma^2}
\end{equation}
for a continuously decoupled qubit. In addition, the argument of $\overline{\langle \tilde \sigma_- \rangle (t)}$ reveals a phase shift that can be interpreted as a small correction to the Rabi frequency, $\Omega\rightarrow\Omega +1/T_2^\Omega$. 

\subsection{Effective qubit-qubit interactions}\label{subsec:effqubitqubit}
\begin{figure}
	\includegraphics[width=\columnwidth]{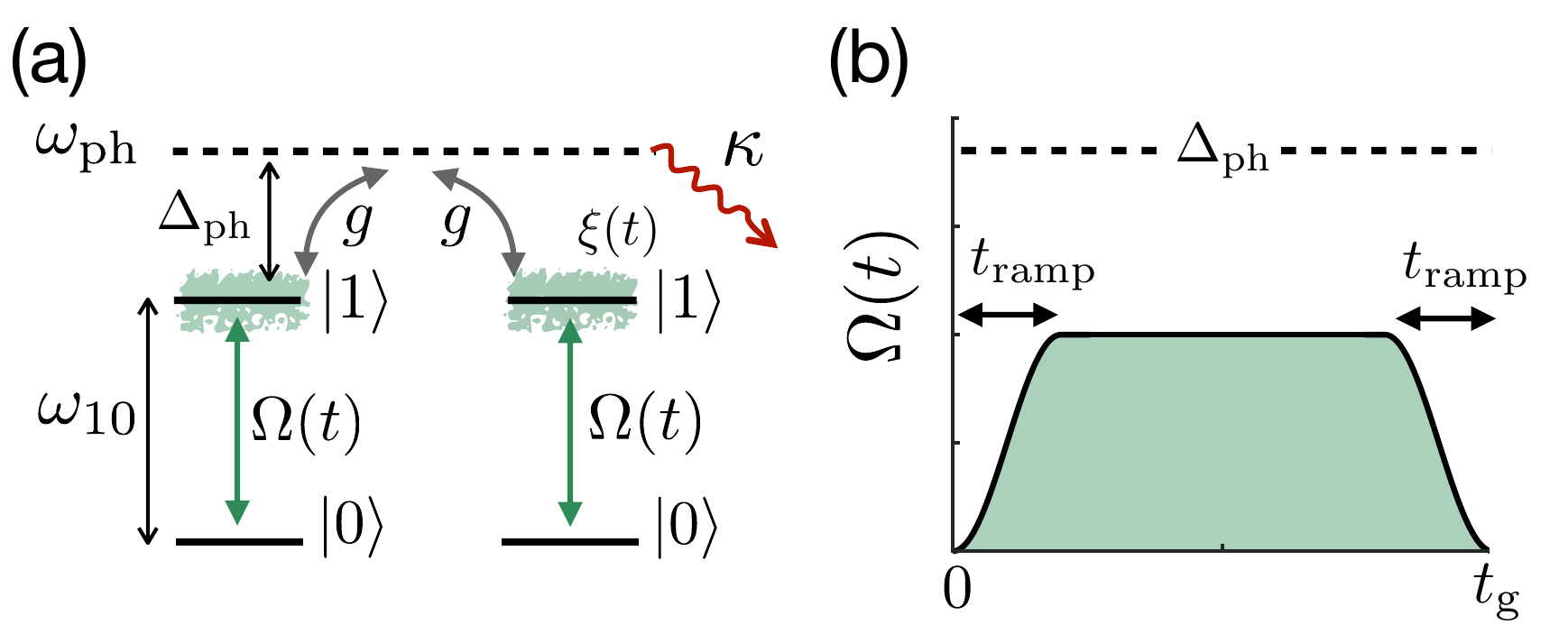} 
	\caption{Phonon-mediated spin-spin interactions. (a) Level scheme for two dressed qubits dispersively coupled to a cavity mode. (b) Sketch of the Rabi frequency $\Omega(t)$ during the implementation of a two-qubit gate operation.  At the beginning (end) of the pulse the Rabi-frequency is ramped up (down) over a time $t_{\rm ramp}$, where $t_{\rm ramp}\ll t_{\rm g}$ in all the numerical examples.} 
	\label{Fig:SpinSpinInteractions}
\end{figure}

The change from the bare to the dressed qubit states not only suppresses the noise, but also modifies the interaction with the phonon mode. By assuming identical frequencies and driving parameters for both quits and after moving into the interaction picture with respect to $H_0=\tilde \Omega \sum_i \tilde \sigma_i^z /2 +  \Delta_{\rm ph} a^\dag a$, the strain coupling Hamiltonian in the dressed qubit basis reads
\begin{equation}\label{Eq:H_full_1mode_Dressed_IP}
\begin{split}
H_{I,\rm e-ph}(t)  =\sum_{i=1}^Ng_i \left[ a^\dag  e^{i\Delta_{\rm ph} t} \Sigma_i (t)  + {\rm H.c.} \right],
\end{split}
\end{equation}
where
\begin{equation}
\Sigma_i(t) = \frac{\sin{(\theta)}}{2}\tilde \sigma_i^z -\sin^2\!{(\theta/2)}\tilde \sigma_i^+e^{i\tilde \Omega t}+\cos^2\!{(\theta/2)} \tilde \sigma_i^-e^{-i\tilde \Omega t}.
\end{equation}
In the limit $\Omega\ll |\Delta|$ we recover the usual Jaynes-Cummings interaction. However, for maximally protected qubits, $\Delta=0$, the coupling to the phonon mode involves additional longitudinal and anti-Jaynes-Cummings terms. As long as all these contributions are off-resonant, i.e., $g_i\ll |\Delta_{\rm ph}|, |\Delta_{\rm ph}\pm \tilde \Omega|$, we can treat them in perturbation theory and derive the effective Hamiltonian [see Fig.~\ref{Fig:SpinSpinInteractions} (a) and Appendix~\ref{app:SWTrans}]
\begin{equation}\label{Eq:H_effective}
\begin{split}
H_{I, \rm eff}=& -J_{\perp}(\tilde\sigma_1^+\tilde\sigma_2^-+\tilde\sigma_1^-\tilde\sigma_2^+) - \frac{J_{\parallel}}{2}\tilde\sigma^z_1\tilde\sigma_2^z \\
&-\sum_{i=1}^2 \Lambda_i(a^\dagger a+1/2)\tilde\sigma_i^z.
\end{split}
\end{equation}
Here we have introduced the spin-spin couplings
\begin{eqnarray}\label{Eq:spinspincouplings}
J_{\perp}&=&g_1g_2\Big\{ \frac{\cos^4(\theta/2)}{\Delta_{\rm ph}-\Omega}+\frac{\sin^4(\theta/2)}{\Delta_{\rm ph}+\Omega} \Big\}, \\
J_\parallel &=&\frac{g_1g_2\sin^2(\theta)}{\Delta_{\rm ph}},
\end{eqnarray}
and
\begin{equation}\label{Eq:Starkcoupling}
\Lambda_i=g_i^2\left[\frac{\cos^4(\theta/2)}{\Delta_{\rm ph}-\Omega}-\frac{\sin^4(\theta/2)}{\Delta_{\rm ph}+\Omega} \right]
\end{equation} 
is the strength of the Stark shift. This effective model shows that the use of dressed spin qubits is fully compatible with the implementation of phonon-mediated gates. Moreover, this coupling scheme is flexible enough, such that even for maximal protection, $\Delta=0$, the ratio between $\Omega$ and $\Delta_{\rm ph}$ can be varied to select between XY-type ($J_\perp\gg J_\parallel$), ZZ-type ($J_\perp\ll J_\parallel$) and fully symmetric ($J_\perp=J_\parallel$) Heisenberg interactions. See also Ref.~\cite{Cao17,Srinivasa23} for closely related proposals for realizing spin-spin interactions by combining dressed qubits with the Jaynes-Cummings model.

\subsection{Effective decoherence sources}
While the use of dressed qubit states suppresses the effect of low frequency noise down to a residual spin coherence time of about $T_2^\Omega$, the small admixing between the qubits and the lossy phonon mode introduces an additional collective decoherence channel for the qubits. As shown in Appendix~\ref{app:SWTrans}, this decoherence mechanism can be described in terms of an effective master equation for the qubits, 
\begin{equation}\label{eq:PID}
\dot{\rho}=\sum_{\mu=+,-,z} \sum_{i,j}\Gamma_{ij}^{\mu} \mathcal{D}_{i,j}[\tilde{\sigma}_\mu],
\end{equation}
where $\mathcal{D}_{i,j}[c]\rho=c_i\rho c_j^\dagger-\frac{1}{2}\{c_i^\dagger c_j,\rho\}_+$. The corresponding rates are
\begin{eqnarray}
\Gamma_{ij}^+\!= g_ig_j\kappa\!\left[ (n_{\rm th}\!+1)\frac{\cos^4(\theta/2)}{(\Delta_{\rm ph}\!-\!\Omega)^2} + n_{\rm th}\frac{\sin^4(\theta/2)}{(\Delta_{\rm ph}\!+\!\Omega)^2} \right], \\
\Gamma_{ij}^-\!= g_ig_j\kappa\!\left[ n_{\rm th}\frac{\cos^4(\theta/2)}{(\Delta_{\rm ph}\!-\!\Omega)^2} + (n_{\rm th}\!+1)\frac{\sin^4(\theta/2)}{(\Delta_{\rm ph}\!+\!\Omega)^2} \right],
\end{eqnarray}
and
\begin{equation}
\Gamma_{ij}^z= \frac{g_ig_j\kappa}{4\Delta_{\rm ph}^2}(2n_{\rm th}+1)\sin^2\theta.
\end{equation}

In addition, as can be seen from the last term of $H_{I,{\rm eff}}$ in Eq.~\eqref{Eq:H_effective}, the phonon mode also induces a Stark shift proportional to the phonon occupation number $a^\dag a$. For any finite temperature, $n_{\rm th}>0$, this shift takes a random value in each experimental run and fluctuates on a timescale $\kappa^{-1}$. In the following, we discuss in more detail how these residual imperfections limit the implementation of gate operations and under which conditions the highest fidelities can be achieved.

\pr{Note that the effective model given in Eq.~\eqref{Eq:H_effective} and Eq.~\eqref{eq:PID} is based on coupling with a single mechanical mode, which we justify by the large mode spacing of about $\sim$ 1 GHz in a micrometer-sized diamond structure. However, all derivations in this section can be easily generalised to multiple phonon modes with frequencies $\omega_{\rm ph}^n$, leading to corrections $\sim 1/(\omega_{10}-\omega_{\rm ph}^n)$ for the effective spin-spin coupling and corrections of $\sim 1/(\omega_{10}-\omega_{\rm ph}^n)^2$ for the effective decay rates. Provided that no random resonances occur, such a multimode generalisation would not significantly change the model parameters or the ratio between coherent and incoherent couplings. Therefore, we continue with the single-mode model for now, but consider the case of a multiphonon lattice system in Sec.~\ref{sec:scalability}.}

\section{High-fidelity entanglement generation}\label{sec:Gates}

We start by assuming $g_1=g_2$ for simplicity and set $\Delta=0$ ($\theta=\pi/2$) for maximal protection. According to our effective model in Eq.~\eqref{Eq:H_effective}, we define the ideal gate Hamiltonian
\begin{equation}\label{Eq:gateHamil} 
H_{\rm g}= -J_{\perp}(\tilde\sigma_1^+\tilde\sigma_2^-+\tilde\sigma_1^-\tilde\sigma_2^+) - \frac{J_{\parallel}}{2}\tilde\sigma^z_1\tilde\sigma_2^z -\frac{\delta \Omega}{2}\sum_{i=1}^2\tilde\sigma_i^z,
\end{equation}
which includes the expected average frequency shift $\delta\Omega=\Lambda(2n_{\rm th}+1)- 1/T_2^\Omega$. For a gate time $t_{\rm g}=\pi/(4J_\perp)$ and up to a global phase, the resulting evolution operator reads
\begin{equation} \label{UnitaryOp}
U_{\rm g}=
\begin{pmatrix}
   e^{i(\gamma +\beta)} & 0 & 0 &  0\\
  0 & 1/\sqrt{2} &  i/\sqrt{2} & 0 \\
   0 &  i/\sqrt{2} &1/\sqrt{2}  & 0 \\
0 & 0 & 0 &  e^{i(\gamma-{\beta})}
\end{pmatrix},
\end{equation}
where $\gamma=\pi J_\parallel/(4J_\perp)$ and ${\beta}=\pi \delta\Omega/(4J_\perp)$. This operation equals the product of an $\sqrt{i \rm SWAP}$ and a C-PHASE gate, which is also known as the fSim gate~\cite{Foxen20}. 

In the presence of noise, but also due to small  inaccuracies of our effective model, the actual evolution of the two-qubit system will slightly deviate from this ideal evolution. In the following we aim to quantify and minimize these residual errors under realistic experimental conditions.

\subsection{Preparation of a Bell state}\label{subsec:Entanglement}
In a first step, we are interested in the generation of a maximally entangled Bell state $|\tilde\Psi\rangle=\frac{1}{\sqrt{2}}(|\tilde{1}\tilde{0}\rangle +i |\tilde{0}\tilde{1}\rangle)$, which is generated by $U_{\rm g}$ when starting from the initial product state $|\tilde 1 \tilde 0\rangle $. To evaluate the expected error of this operation in a realistic system, we define the state fidelity $\mathcal{F}=\langle \tilde \Psi | \rho_q(t_{\rm g})|  \tilde \Psi \rangle$, where $\rho_{q}(t_{\rm g})={\rm Tr}_{\rm ph}\{\rho(t_{\rm g})\}$ is the reduced qubit state at the end of the gate. To facilitate an analytical treatment, we assume the error $\mathcal{E}=1-\mathcal{F}\ll1$ to be small. Then, using perturbation theory, we can write the total gate error in the form (see Appendices~\ref{app:analyticerrors} and~\ref{app:SPEerror})
\begin{equation}\label{Eq:Gate_error}
\mathcal{E}_{\Psi}= \mathcal{E}_\varphi + \mathcal{E}_\kappa+\mathcal{E}^{\rm min}_{\rm s-ph}.
\end{equation}
Here, the first term,
\begin{equation}
\mathcal{E}_\varphi = \frac{1}{2}\Big(\frac{t_{\rm g}}{T_2^\Omega}\Big)^2,
\end{equation}
represents the residual error caused by the original low-frequency noise. The second term
\begin{equation}
\mathcal{E}_\kappa=\frac{1}{2} \left(\Gamma^+_{11}+\Gamma^+_{22}+ \Gamma^-_{11}+\Gamma^-_{22}\right)t_{\rm g},
\end{equation}
accounts for the phonon-induced decoherence described by Eq.~\eqref{eq:PID}. Finally, there is a third source of error, $\mathcal{E}^{\rm min}_{\rm s-ph}$, which arises from the fact that at the end of the gate, the qubits are still slightly entangled with the phonon mode. By switching the Rabi frequency on and off adiabatically at the beginning and end of the gate, this error can be reduced to 
\begin{equation}
\mathcal{E}^{\rm min}_{\rm s-ph}\simeq (2n_{\rm th}+1)\frac{g^2}{\Delta_{\rm ph}^2},
\end{equation}
but not completely avoided. Note that in all our numerical simulations below, we assume an adiabatic ramp $\Omega(t)=\Omega\sin^2[\pi (t-t_m)/(2t_{\rm ramp})]$ at the beginning ($t_m=0$) and at the end ($t_m=t_{\rm g}$) of the gate [see Fig.~\ref{Fig:SpinSpinInteractions} (b)] and optimize the ramp duration $t_{\rm ramp}\ll t_{\rm g}$. The gate errors are evaluated using the reduced qubit state after the second ramp.

\subsection{Parameter optimization}\label{subsec:Optimization}
\begin{figure}
	\includegraphics[width=\columnwidth]{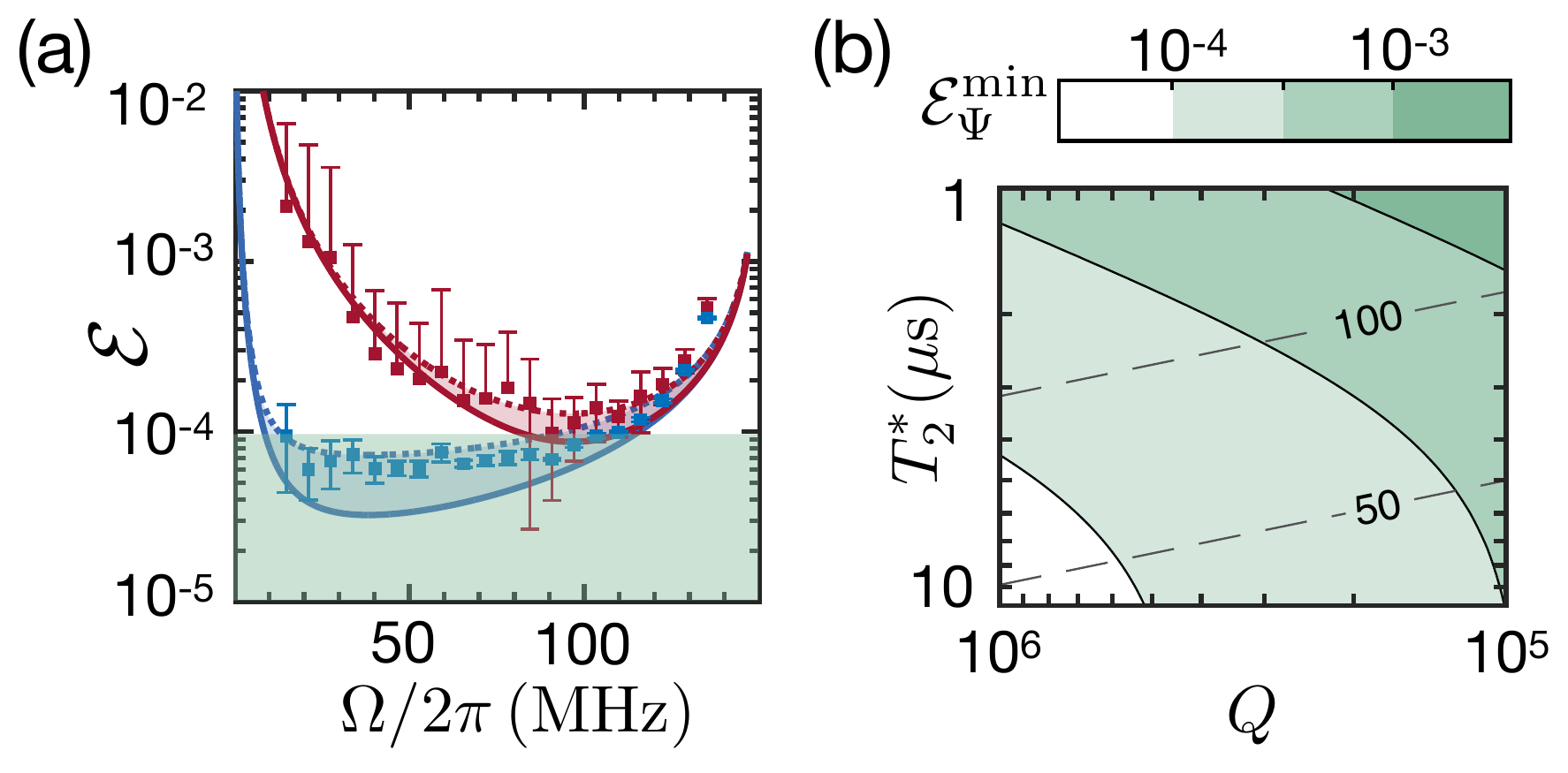} 
	\caption{
 Preparation of a Bell state. (a) Plot of the residual error $\mathcal{E}$ versus the Rabi frequency $\Omega$, for the initial two-qubit state $|\tilde{\Psi}\rangle$. The dotted (solid) lines represent Eq.~\eqref{Eq:Gate_error} including (without) the contribution $\mathcal{E}_{\rm s-ph}^{\rm min}$ for $T_2^*=10 \mu$s (blue) and $3 \mu$s (red). The markers indicating the result of exact master equation simulations fall between the two lines, showing a good correspondence between analytic and numerical results. Each point is the result of 100 independent noise realisations, where the $\xi_i$ were randomly selected from the probability distribution $P(\xi)=(2\pi\sigma^2)^{-1/2}\exp{(-\xi^2/2\sigma^2)}$. Other relevant parameters are $Q=10^6$, $T=100$ mK, $g/(2\pi)\approx 0.75$ MHz and $\Delta_{\rm ph}/(2\pi)\approx 150$ MHz. (b) Plot of the minimal error $\mathcal{E}_\Psi^{\rm min}$ in Eq.~(\ref{Eq:Min_Gate_error}) versus $Q$ and $T_2^*$. 
 The two dashed lines indicate the boundary in this parameter space above which $\Omega_{\rm opt}/(2\pi)$ surpasses the values of $50$ MHz and $100$ MHz, respectively. This becomes relevant when there are additional experimental constraints on the maximal value of $\Omega$.} 
	\label{Fig:EntError}
\end{figure}

Figure~\ref{Fig:EntError} (a) shows examples of the simulated errors for the entanglement generation operation as a function of the Rabi frequency $\Omega$ and for two different values of $T_2^*$. In this plot the square markers represent the results obtained from an exact simulation of the full master equation in Eq.~\eqref{Eq:Master_Equation}, starting from the initial state $\rho_{\rm ini}=|\tilde 1 \tilde 0\rangle\langle \tilde 1 \tilde 0|\otimes \rho_{\rm th}$ and averaged over many noise realizations. The solid and dotted lines correspond to the analytic formula in Eq.~\eqref{Eq:Gate_error}, including and without including the contribution of $\mathcal{E}^{\rm min}_{\rm s-ph}$. While even for a highly coherent system it is not always possible to reach errors as low as $10^{-4}$, this threshold is achievable within a certain parameter range and under optimized driving conditions. 

To identify these conditions we can use the approximate analytic expression for $\mathcal{E}_\Psi$ given in Eq.~\eqref{Eq:Gate_error}, which is in excellent agreement with the exact numerical results. From this expression, it first follows that we must choose a minimal phonon detuning of about $\Delta_{\rm ph} \gtrsim 10^{2} g \sqrt{2n_{\rm th}+1}$ in order to suppress the residual spin-phonon entanglement to a level of $\mathcal{E}^{\rm min}_{\rm s-ph}<10^{-4}$. Once this detuning is fixed, the remaining error, $\mathcal{E}_\varphi+\mathcal{E}_\kappa$, can be minimized by choosing an optimal value $\Omega_{\rm opt}\simeq \Delta_{\rm ph}/(1+r)$ for the Rabi frequency (see Appendix~\ref{app:Minimization}). Here $r=\frac{1}{2}(\mathcal{E}_\kappa^{\min}/\mathcal{E}_\varphi^{ 0})^{1/3}$ with $\mathcal{E}_\kappa^{\min}=\pi\kappa(2n_{\rm th}+1)/(4\Delta_{\rm ph})$ and $\mathcal{E}_\varphi^{ 0}=\frac{\pi^2}{8}(gT_2^*)^{-4}$.  For a given $r$, the resulting expression for the minimal total error is 
\begin{equation}\label{Eq:Min_Gate_error}
\mathcal{E}_{\Psi}^{\rm min}= \left(\frac{\pi}{4}\right)^{\frac{4}{3}}\frac{G(r)}{[\mathcal{C}_\Omega (1+r)]^{2/3}}  + \mathcal{E}^{\rm min}_{\rm s-ph},
\end{equation}
with $G(r) = \frac{(1+r)^2+1}{(2+r) } + \frac{1}{2^3} \frac{(2+r)^2}{(1+r)^2}$. Here we have introduced the effective cooperativity \begin{equation}
\mathcal{C}_\Omega=(\Omega_{\rm opt}T_2^*/4)\mathcal{C}^*,
\end{equation}
which is enhanced by the Rabi frequency. For small $r$ we then obtain the error scaling 
\begin{equation}\label{eq:Error_Min}
\mathcal{E}_{\Psi}^{\rm min}\approx  \frac{3}{2}\left(\frac{\pi}{4 \sqrt{\mathcal{C}_\Omega}}\right)^{\frac{4}{3}} \sim \mathcal{C}_\Omega^{-\frac{2}{3}},
\end{equation}
which is similar to the scaling for unprotected qubits, but with an enhanced cooperativity that is boosted by a factor $\Omega_{\rm opt} T_2^*/4$. Note, however, that this limit is in practice not always attainable, for example, due to constraints on the maximal Rabi-frequency $\Omega_{\rm opt}\sim \Delta_{\rm ph}$. In this case the minimally achievable error will be dominated by phonon losses and spin-phonon entanglement, i.e., 
\begin{equation}
\mathcal{E}_{\Psi}^{\rm min}\approx \mathcal{E}_\kappa^{\min}+\mathcal{E}^{\rm min}_{\rm s-ph}.
\end{equation}

As  a specific example, we consider a phonon mode with $\omega_{\rm ph}/(2\pi)\approx 3$ GHz, $Q=10^6$, and $T=100$ mK (${n}_{\rm th}\approx0.3$), $T_2^*=10 \,\mu$s and $g/(2\pi)\approx 0.75$ MHz. For a detuning of  $\Delta_{\rm ph}/(2\pi)\approx 150$ MHz we then obtain $\mathcal{E}_{\Psi}^{\rm min}\approx 7 \times 10^{-5}$, with $\Omega_{\rm opt}/(2\pi)\approx 45$ MHz and a gate time $t_{\rm g}\approx 60\, \mu$s.  For the same coupling strength, we plot in Fig.~\ref{Fig:EntError} (b) the resulting minimal gate error as a function of the mechanical Q-factor and the bare spin coherence time. While we see that reaching a level of $10^{-4}$ is challenging, the required parameters are well within reach of near-term experiments. 

\section{Universal two-qubit gates}\label{sec:HFgates}
In the previous example, the evolution from the initial state to the final entangled state was constraint to the subspace $\{|\tilde{0}\tilde{1}\rangle,|\tilde{1}\tilde{0}\rangle\}$. For  $g_1=g_2$, these states are insensitive to the common Stark shift appearing in Hamiltonian~\eqref{Eq:H_effective}. Therefore this term does not contribute to the error budget in Eq.~\eqref{Eq:Gate_error}. However, this is not true for the states $\{|\tilde{0}\tilde{0}\rangle,|\tilde{1}\tilde{1}\rangle\}$, or, more generally, when $g_1\neq g_2$. To assess the fidelity of a universal two-qubit gate independently  of the input states, the additional decoherence channel arising from a finite thermal population of the resonator mode must be taken into account.  

\subsection{Thermal Stark-shift dephasing}\label{subsec:ThermalDephasing}
Since the average thermal frequency shift has already been taken into account in our target gate Hamiltonian in Eq.~\eqref{Eq:gateHamil}, thermal decoherence arises from the remaining fluctuating contribution, 
\begin{equation}\label{eq:HStarkfluc}
H_{\rm Stark}= \Lambda (a^\dag a - n_{\rm th}) (\tilde \sigma^z_1+\tilde \sigma^z_2),
\end{equation}
assuming $\Lambda_i=\Lambda$. 
In order to determine an upper bound for the influence of this shift on the implementation of universal two-qubit gates, we consider the same evolution as above, but starting from the initial state $|\tilde \Phi\rangle=\frac{1}{\sqrt{2}}(|\tilde{1}\tilde{1}\rangle +i |\tilde{0}\tilde{0}\rangle)$. For this state,  which is maximally sensitive to this shift, we obtain an additional contribution to the total gate error, which is of the form (see Appendix~\ref{app:analyticerrors})  
\begin{equation}\label{Eq:StarkError}
\mathcal{E}_{\Lambda}\approx  n_{\rm th}(n_{\rm th}+1) \frac{4\Lambda^2}{\kappa^2}  [e^{-\kappa t_{\rm g}}-(1-\kappa t_{\rm g})].
\end{equation} 
For $\kappa t_{\rm g}\ll1$, we obtain the scaling $\mathcal{E}_{\Lambda}\sim \Lambda^2 t_{\rm g}^2$, which reflects the static uncertainty in the initial phonon occupation number. In the opposite regime, $\kappa t_{\rm g}\gg1$, the number fluctuations are correlated only for a short time and the resulting dephasing error, $\mathcal{E}_{\Lambda}\sim (\Lambda^2/\kappa) t_{\rm g}$, scales more favorably. Note that for the parameters considered in Fig.~\ref{Fig:EntError} (a), $\kappa t_{\rm g}\approx 1$ and we are therefore between these two limits. Importantly, the dephasing is thermally activated and is strongly suppressed for $n_{\rm th}\ll1$. Therefore, this mechanism is usually not considered in optical cavity QED systems, where the thermal population of the cavity mode is negligible. For phononic cavities, this is not necessarily the case.

\begin{figure}
\includegraphics[width=\columnwidth]{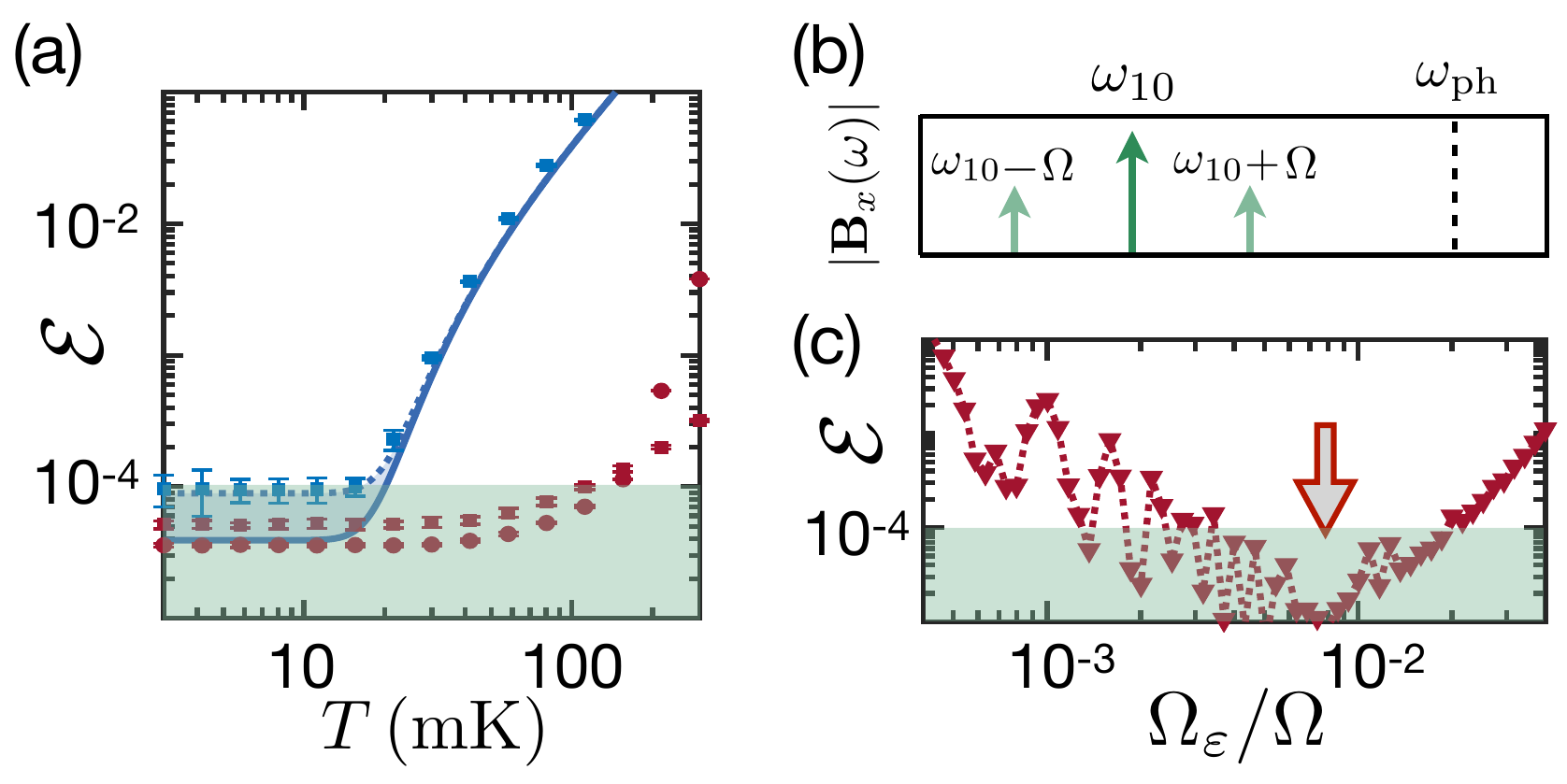} 
\caption{High-fidelity two-qubit gate. (a) Plot of the error $\mathcal{E}$ versus the phonon temperature $T$. The blue dotted (solid) line represents $\mathcal{E}_{\Phi}$ including (without including) the contribution from $\mathcal{E}_{\rm s-ph}^{\rm min}$. The blue markers correspond to the results obtained from exact master equation simulations with an initial two-qubit state $|\tilde{\Phi}\rangle$, averaged over 50 random static-noise realizations. The red square (round) markers are the results from exact numerical simulations for the initial state $|\Phi\rangle$ ($|\Psi\rangle$), and using the CCDD method. (b) Sketch of the relevant frequency components of ${\bf B}_{x}(t)$ in the CCDD scheme. In addition to the central driving frequency at $\omega_{10}$, the phase modulation generates two weaker sidebands at $\omega_{10}\pm\Omega$. (c) Plot of the gate error $\mathcal{E}$ for the CCDD scheme as a function of $\Omega_\varepsilon$ and for an initial state $|\Phi\rangle$. Note that for this plot only the Hamiltonian evolution has been taken into account and for each value of $\Omega_\varepsilon$, $t_{\rm ramp}$ has been slightly adjusted to minimize the error. The dotted line is a guide to the eye. In (a), $\Omega_{\varepsilon}/\Omega=8\times10^{-3}$ and $t_{\rm ramp}=2.15\times 2\pi/\Delta_{\rm ph}$ were chosen based on the most favorable values identified in (c). The other relevant parameters are $T_2^*=10 \,\mu$s, $Q=10^6$, $g/(2\pi)=0.75$ MHz, $\Delta_{\rm ph}/(2\pi)\approx 150$ MHz and $\Omega/(2\pi)\approx 37 $ MHz.} 
	\label{Fig:CCDD}
\end{figure}

In Fig.~\ref{Fig:CCDD} (a) we simulate the implementation of the two-qubit gate $U_{\rm g}$ in Eq.~\eqref{UnitaryOp} for the initial state $|\tilde\Phi\rangle$ and plot the total error of this operation as a function of the temperature $T$. The blue solid curve represents the analytic prediction 
\begin{equation}
\mathcal{E}_{\Phi} = \mathcal{E}_{\Psi} + \mathcal{E}_{\Lambda},
\end{equation}
while the blue markers are the results of exact numerical simulations. For $T=100$ mK, which corresponds to $n_{\rm th}\approx 0.3$ in the current example, the error is well above $10^{-2}$ and dominates over all other contributions discussed above. Since in practice it might be difficult to cool the sample significantly below this temperature, these simulations show that thermal frequency shifts are a major limitation for realizing phonon-mediated gate operations with sufficiently high fidelity.

\subsection{Concatenated decoupling}\label{subsec:CCDD}

To overcome this problem,  we propose to use a method called concatenated continuous dynamical decoupling (CCDD)~\cite{Cai12,Cohen16,Farfurnik17,Stark17,Wang20,Cao20,Ramsay23,Wang2023,Martinez2024}, which is known to effectively improve the coherence times of dressed qubits affected by low-frequency fluctuations of the Rabi frequency.  The basic idea behind this method  is to replace the constant microwave field by a phase-modulated field,
\begin{equation}
{\bf B}_x= \mathcal{B} \cos\left[\omega_{10} t-2\varepsilon \sin(\Omega t)\right]. 
\end{equation}
By assuming that $\varepsilon\ll 1$ and moving into a rotating frame with respect to $\omega_{10}$, the bare qubit Hamiltonian then reads
\begin{equation}\label{CCDDHamil}
H_q = \frac{\Omega}{2} \sigma_x -\Omega_\varepsilon \sin(\Omega t)\sigma_y,
\end{equation}
where $\Omega_\varepsilon=\varepsilon\Omega$. We see that the dominant energy splitting of the dressed states is still determined by $\Omega$ and small  quasi-static shifts $\sim\sigma_z$  are suppressed as discussed above. The second term $\sim\sigma_y$ is modulated with exactly this transition frequency and induces oscillations between the dressed states $|\tilde 0\rangle$ and $|\tilde 1\rangle$ with Rabi frequency $\Omega_\varepsilon$. We can therefore repeat the idea of continuous dynamical decoupling and use a sufficiently large $\Omega_\varepsilon$ to suppress small quasi-static contributions $\sim\delta\Omega \sigma_x$, which correspond to a term $\sim \delta\Omega \tilde \sigma_z$ in the dressed-state basis. 
Since the Stark-shift Hamiltonian in Eq.~\eqref{eq:HStarkfluc} is of that form, we expect the CCDD method to be applicable in this case as well. 

Let us now combine these ideas with phonon-mediated spin-spin-interactions. In the following, we restrict ourselves to the case $\Delta=0$ and assume that $\Omega_\varepsilon\ll \Omega, \Delta_{\rm ph}$ does not considerably change the qubit-phonon detunings. In this case we can simply repeat the adiabatic elimination of the phonon mode and obtain the following effective spin model in the interaction picture
\begin{equation}
\begin{split}
H_{I,{\rm eff}}(t)  \simeq \frac{\Omega_\varepsilon}{2}\sum_i\tilde{\sigma}_i^x-\sum_i \Lambda_i(a^\dagger a+1/2)\tilde{\sigma}_i^z \\
-J_{\perp}(\tilde{\sigma}_1^+\tilde{\sigma}_2^-+\tilde{\sigma}_1^-\tilde{\sigma}_2^+) - \frac{J_{\parallel}}{2}\tilde{\sigma}^z_1\tilde{\sigma}_2^z.
\end{split}
\end{equation}
We see that, compared to Eq.~\eqref{Eq:H_effective}, we have an additional local driving term that we use to suppress the Stark-shift fluctuations. To do so, we make another unitary transformation into an interaction picture with respect to $H_0=\frac{\Omega_\varepsilon}{2}\sum_i\tilde{\sigma}_i^x$, and neglect all terms that oscillate with $e^{\pm i\Omega_\varepsilon t}$. This leaves us with the ideal gate Hamiltonian  
\begin{equation}\label{eq:Hg_concat}
H'_{\rm g} \simeq -J'_{\perp}({\sigma}_1^+{\sigma}_2^-+{\sigma}_1^-{\sigma}_2^+) - \frac{J'_{\parallel}}{2}{\sigma}^z_1{\sigma}_2^z,
\end{equation}
where $J'_\perp=(J_\perp+J_\parallel)/2$ and $J'_\parallel=J_\perp$. Note that due to this second basis transformation, the final effective interaction is expressed in terms of the original Pauli operators, but the overall strength of the interaction and the degree of tunability is similar to the non-concatenated driving scheme. 

\subsection{Example}\label{subsec:ExampleCCDD}
Note that in the derivation of Eq.~\eqref{eq:Hg_concat} we implicitly assumed that the additional modulation sidebands $\omega_{10}\pm \Omega\neq \omega_{\rm ph}$ do not accidentally match the phonon frequency, and we have neglected many terms that oscillate at various different frequencies. To justify the validity of all these approximations and to verify that the CCDD scheme is indeed able to suppress thermal fluctuations, we compare in Fig.~\ref{Fig:CCDD} (a) the dynamics generated by $H^\prime_{\rm g}$ with exact numerical simulations of the full system in Eq.~\eqref{Eq:Master_Equation} (see Appendix~\ref{app:CCDD} for further details). The square red markers represent the results of the numerical simulations for the initial state $|\Phi\rangle=\frac{1}{\sqrt{2}}(|{1}{1}\rangle +i |{0}{0}\rangle)$. The round markers represent the result for the initial state $|\Psi\rangle=\frac{1}{\sqrt{2}}(|{1}{0}\rangle +i |{0}{1}\rangle)$. 
Contrary to the gate using only CDD, the gate using the CCDD method achieves errors below $10^{-4}$ even at $T=100$ mK. For the other parameters in this simulation we considered $T_2^*=10 \,\mu$s, $Q=10^6$, $g/(2\pi)=0.75$ MHz, $\Omega/\Delta_{\rm ph}=1/4$ ($\Omega/(2\pi)\approx 37 $ MHz), $\Omega_\varepsilon/\Omega=8\times10^{-3}$ and $t_{\rm ramp}=2.15\times 2\pi/\Delta_{\rm ph}$. The value of $\Omega$ was chosen close to $\Omega_{\rm opt}$, while $\Omega_\varepsilon$ was chosen based on the results from numerical Hamiltonian simulations shown in Fig.~\ref{Fig:CCDD} (c). Simulations starting with other initial states yield similar results. Based on our results, we conclude that two-qubit gates below the error threshold of $10^{-4}$ are realistically achievable with a CCDD scheme.

\section{Scalability}\label{sec:scalability}
In Sec.~\ref{sec:DressedQubits}, we investigated the implementation of  high-fidelity two-qubit gates between two SiV centers that are coupled to a single localized phonon mode, as sketched in Fig.~\ref{Fig:Schema}. The same protocol can in principle be extended to multiple spins coupled to the same mode. However, due to a finite mode volume and the need to address qubits individually, such an approach is not scalable to tens or hundreds of spins. 

\begin{figure*}
	\includegraphics[width=1.7\columnwidth]{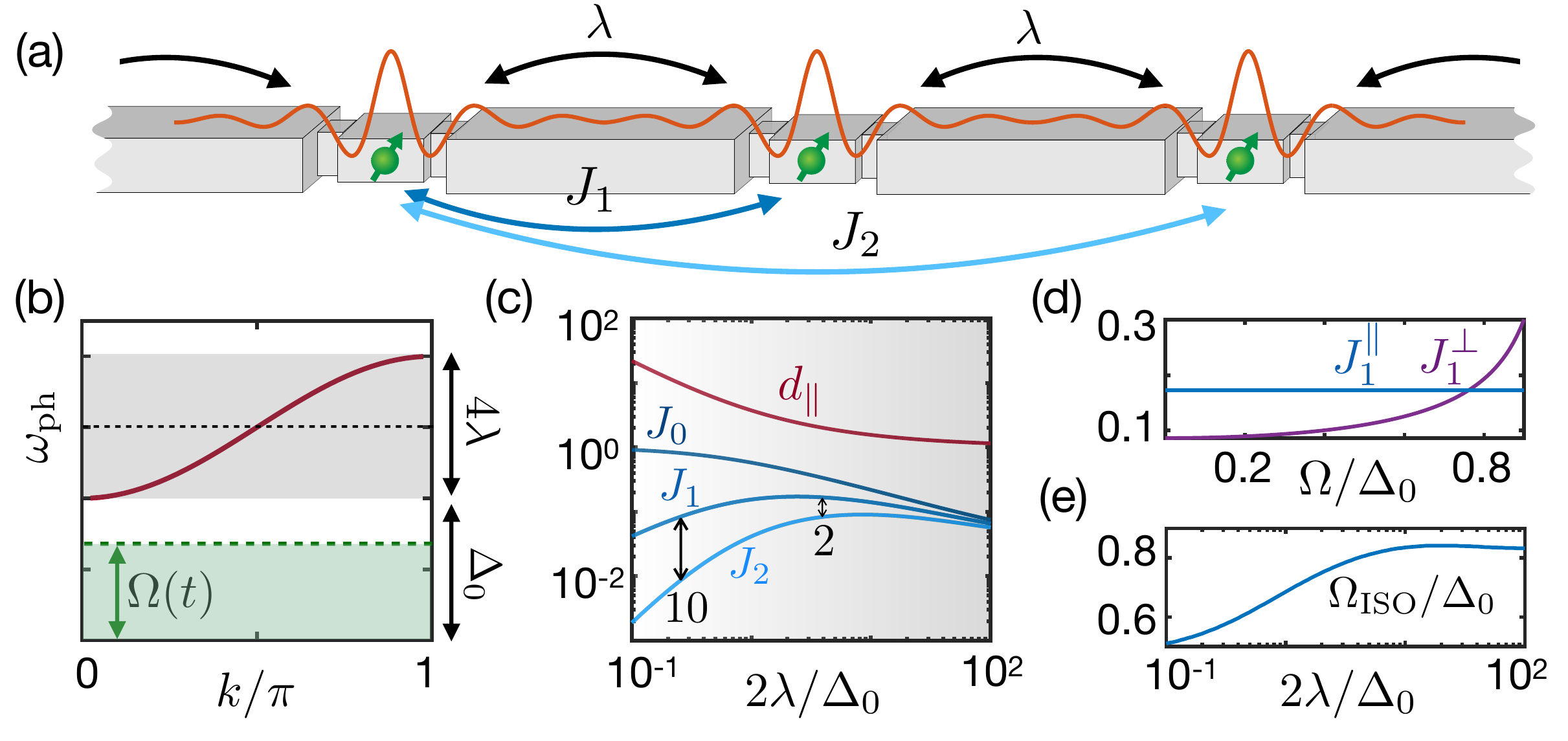} 
 \caption{Effective spin-spin interactions in phononic crystals (a) Schematic representation of SiV centers embedded in a one-dimensional phononic crystal. Each center interacts with a localized phonon mode, ideally at a position where the amplitude of the strain field (indicated by the orange line) is the largest. At the same time, the localized mode couples ($\sim \lambda$) to neighboring modes via a small evanescent overlap between the strain profiles. This leads to the formation of extended Bloch waves, which mediate spin-spin interactions over a range of lattice sites, controlled by the detuning $\Delta_0$. (b)~Dispersion relation of the phononic modes (solid red), where $4\lambda$ is phononic bandwidth and $\Delta_0$ is the detuning between the $k=0$ mode frequency and the qubit frequency. (c) Characteristic length $d_\parallel$ and spin-spin couplings $J_{n=1,2,3}\equiv J^\parallel_{i,i+n}/J_{i,i}$ versus $2\lambda/\Delta_0$. The nearest-neighbour coupling $J_1$ achieves its largest value ($J_1\approx 0.17$) at $2\lambda/\Delta_0=2.42$. (d) Characteristic nearest-neighbour couplings $J^\parallel_{1}\equiv J^\parallel_{i,i+1}/J_{i,i}$ and $J^\perp_{1}\equiv J^\perp_{i,i+1}/J_{i,i}$ for different values of the Rabi frequency $\Omega$ and for $2\lambda/\Delta_0=2.42$. (e) Value of the Rabi frequency $\Omega_{\rm ISO}$, for which $J^\perp_{1}$ equals $J^\parallel_{1}$, versus the tunneling rate to the neighboring lattice sites $\lambda$.} 
	\label{Fig:Scalability}
\end{figure*}

\subsection{Spin-phonon superlattices}

As a way to scale-up this spin-phonon platform, one can make use of multiple coupled phononic cavities within the crystal structure. Fig.~\ref{Fig:Scalability} (a) shows a sketch of such a phononic crystal, where multiple localized phonon modes are coupled to neighboring modes via a small evanescent overlap between the strain profiles. Note that more specific designs for such coupled phononic cavity arrays have been discussed, for example, in Refs.~\cite{Kuzyk18,Li19}. For the current analysis, we simply assume that this coupled phonon array is well-described by a tight-biding lattice Hamiltonian with a tunneling amplitude $\lambda\ll\omega_{\rm ph}$ and that each cavity hosts only a single SiV spin with the same type of spin-phonon coupling introduced in Sec.~\ref{sec:Model}. The resulting Hamiltonian for the whole lattice with $N$ sites reads
\begin{equation}\label{eq:SpinPhononLattice}
\begin{split}
H = \omega_{\rm ph} \sum_{j=1}^{N} a_j^\dagger a_j - \lambda\sum_{j=1}^{N-1}(a_ja^\dagger_{j+1}+a^{\dagger}_ja_{j+1}) \\  +\sum_{j=1}^{N} \frac{\omega_{10}}{2}\sigma_j^z  +\sum_{j=1}^{N} \Omega_j\cos{(\omega_j t-\phi_j)}\sigma_j^x \\  +\sum_{j=1}^N g_j\left( a_j^\dag \sigma^-_j + a_j \sigma_j^+\right).
\end{split}
\end{equation}
In a normal-mode or Bloch-state basis the first line in Eq.~\eqref{eq:SpinPhononLattice} becomes $H_{\rm ph}=\sum_{k=k_{\rm min}}^{k_{\rm max}} \omega^{\rm ph}_k c_k^\dagger c_k$. Here, 
\begin{equation}
\omega^{\rm ph}_k=\omega_{\rm ph}-2\lambda\cos{(k)}
\end{equation}
and $c_k=\sum_{j} b_{jk}a_j$, with $b_{jk}=\sqrt{\frac{2}{N+1}}\sin{(k j)}$ and $k=\pi m/(N+1)$ with $m\in\mathbb{N}$ going from $1$ to $N$ for open boundary conditions. Also, from now on, $\sum_k \equiv\sum_{k=k_{\rm min}}^{k_{\rm max}}$. On resonance, $\omega_j=\omega_{10}$, in a rotating frame with respect to $\sum_k \omega_{10} c_k^\dagger c_k+\sum_{j=1}^{N} \frac{\omega_{10}}{2}\sigma_j^z$, and after neglecting fast rotating terms, Hamiltonian~\eqref{eq:SpinPhononLattice} reduces to
\begin{equation}\label{DrivenJCMultimode}
\begin{split}
H\simeq&\sum_k \Delta_kc^\dagger_kc_k+ \sum_{j=1}^{N} \frac{\Omega_j}{2}\sigma_j^x \\ &+\sum_{j,k}^Ng_{j} b_{jk} \left(c_k^\dag \sigma^-_j +  c_k \sigma_j^+\right),
\end{split}
\end{equation}
where $\Delta_k=\omega_{\rm ph}-\omega_{10}-2\lambda\cos{k}$. For $N\gg1$, the lowest detuning is given by $\Delta_0= \omega_{\rm ph}-\omega_{10}-2\lambda$. Similar to the case of a single mode discussed above, we assume $|b_{jk}g_j|\ll {\rm min}\{\Delta_0-\Omega_j\}$ to  ensure a purely dispersive coupling between spins and phonons, see Fig.~\ref{Fig:Scalability} (b). 

\subsection{Engineering spin models}
Following the same procedure as in Sec.~\ref{subsec:effqubitqubit}, we can eliminate the phonon modes and obtain the effective Hamiltonian 
\begin{equation}\label{Eq:SpinSpinHam} 
\begin{split}
H_{I, \rm eff}=& \sum_{i<j}- J_{ij}^{\perp}(\tilde{\sigma}_i^+\tilde{\sigma}_j^-+\tilde{\sigma}_i^-\tilde{\sigma}_j^+) - \frac{J_{ij}^{\parallel}}{2}\tilde{\sigma}^z_i\tilde{\sigma}_j^z \\
&+\sum_{j}\frac{\delta_j}{2}\tilde{\sigma}_j^z-\sum_{j,k} \Lambda_{jk}c_k^\dagger c_k\tilde{\sigma}_j^z,
\end{split}
\end{equation}
where
\begin{eqnarray}
J_{ij}^{\perp}&=&\sum_{k} \frac{g_ig_j b_{ik} b_{jk}}{4}\Big[\frac{1}{\Delta_k-\Omega} + \frac{1}{\Delta_k+\Omega}\Big], \\
J_{ij}^{\parallel}&=& \sum_{k}  \frac{g_ig_j b_{ik} b_{jk}}{\Delta_k},\\
\Lambda_{jk}&=&\frac{g_j^2 b^2_{jk}}{4}\Big[\frac{1}{\Delta_k-\Omega} - \frac{1}{\Delta_k+\Omega}\Big],
\end{eqnarray}
and $\delta_j=\Omega_j-\Omega- \sum_{k}\Lambda_{jk}$.

The first line of Eq.~\eqref{Eq:SpinSpinHam} represents a generic spin model, where the relative strength between the transverse ($\sim J^\perp$) and longitudinal ($\sim J^\parallel$) interactions, as well as their range, depend on $\lambda$, $\Delta_0$ and $\Omega$. 
To analyze these dependencies, we take the mode-continuum limit $N\rightarrow \infty $, in which case we obtain
\begin{equation}\label{WaveguideCouplings}
\frac{ J_{ij}^\perp}{J_{ij}} \approx \frac{\alpha_{+} }{4d_+^{|i-j|}}+\frac{\alpha_{-}} {4d_-^{|i-j|}} , \qquad \frac{J_{ij}^\parallel}{J_{ij}}\approx \frac{\alpha_{\parallel}}{d_\parallel^{|i-j|}}.
\end{equation}
Here, $J_{ij}=g_ig_j/(2\Delta_0)$ and 
\begin{equation}
\alpha_{\mu}=\frac{1}{(1+x_\mu)\cos{\theta_\mu}},  \qquad d_{\mu}=\frac{\sin{\theta_\mu}}{1-\cos{\theta_\mu}}
\end{equation}
with $x_{\parallel}=2\lambda/\Delta_0$, $x_{\pm} =x_\parallel \pm\Omega/\Delta_0$, $\sin \theta_\parallel=x_\parallel/(1+x_\parallel)$ and $\sin \theta_\pm=x_\parallel/(1+x_\pm)$. For more details, see Appendix~\ref{app:ContinuumCouplings}.

In Fig.~\ref{Fig:Scalability} (c), we plot the characteristic length $d_\parallel$, which describes the decay length of the interactions,  and  the couplings $J_{n=1,2,3}\equiv J^\parallel_{i,i+n}/J_{i,i}$ as a  function of $2\lambda/\Delta_0$. First, we observe that the range of spin-spin interactions depends strongly on $2\lambda/\Delta_0$. For $2\lambda/\Delta_0\ll1$ we have short-range spin-spin interactions, while for $2\lambda/\Delta_0\gg1$ these are long range. The behaviour for $J^\perp_{i,j}$ is qualitatively similar. In addition, we find that for $2\lambda/\Delta_0=2.42$, the nearest-neighbor coupling-strength achieves the largest value $J^\parallel_{i,i+1}\approx 0.17 J_{i,i}$.  For this value of $\lambda$,  we plot in Fig.~\ref{Fig:Scalability} (d) the values of  $J_{i,i+1}^\perp$ and $J_{i,i+1}^\parallel$ as a function of $\Omega/\Delta_0$. This plot shows that, as in the single-mode case, the ratio between the XY ($J_\perp$) and the longitudinal Ising couplings ($J_\parallel$)  can be tuned over a large parameter range by simply varying the Rabi frequency $\Omega$. Furthermore, this property is found  for any value of $2\lambda/\Delta_0$, such that the range of interactions can be tuned independently.  The latter is illustrated in Fig.~\ref{Fig:Scalability} (e), where the Rabi frequency $\Omega$ for which $J_{i,i+1}^\perp= J_{i,i+1}^\parallel$ is achieved, i.e. $\Omega_{\rm ISO}$, is plotted as a function of $2\lambda/\Delta_0$.

\subsection{Protected spin qubits in a scalable phononic lattice}
The previous discussion shows that even in a phononic lattice systems, the same continuous dynamical decoupling approach can be used to engineer protected spin-spin interactions of varying range and type. When only neighboring spins are addressed, this can in principle be used to implement high-fidelity quantum gates in a scalable quantum register. Alternatively, when coupling all spin qubits simultaneously,  the same setup realizes a flexible quantum simulator for Heisenberg-type spin models. Note that for both applications, a generalization to two-dimensional lattice systems is possible.   

While for a large phononic lattice it is no longer possible to simulate the exact dynamics, we argue that the transition to a multi-mode phononic crystal setting does not significantly degrade the high-fidelity operations predicted for the single-mode case. First of all, the continuous dynamical protection mechanism does not change due to the transition from a single mode to a multi-mode system, and all the results presented in Fig.~\ref{Fig:Scalability} have been obtained for maximally protected qubits with $\Delta=0$. Therefore, the suppression of local low-frequency noise can be ensured independently of the engineered spin model. Also, under the same conditions (with $\Delta_0$ replacing $\Delta_{\rm ph}$), the errors related to a finite admixture of the lossy phonon modes will be similar or even reduced compared to the single-mode case, since for $\lambda \gtrsim \Delta_0$ the hybridization with the higher-frequency modes is suppressed. Note however, that the effective coupling strength between nearest neighbour spins is slightly smaller than in the single-mode case [see Fig.~\ref{Fig:Scalability} (c)], which overall might require a slightly higher spin phonon coupling strength $g$ to obtain the same gate fidelities. 

Finally, let us also address the effect of thermal Stark shifts, which we have identified above as a major source of error. In the single mode case, we see from Eq.~\eqref{Eq:StarkError} that this error is caused by thermal fluctuations of the mode occupation number, which are correlated over a timescale $\kappa^{-1}$. In Appendix~\ref{app:ContinuumCouplings} we repeat the same analysis for the case of a single spin qubit coupled to an infinite phononic lattice and find that the corresponding Stark-shift dephasing error $\mathcal{E}_\Lambda$ is reduced by a factor of about $(\kappa/\pi\lambda)\times 100\approx 10^{-3}$. Apart from the numerical factors, this reduction can be explained by the fact that in a lattice system the relevant correlation time for local thermal fluctuations is set by the tunneling rate to the neighboring lattice sites, $\lambda$, and not the phonon loss rate $\kappa$. Therefore, thermal fluctuations affect the qubit frequency for a much shorter time and the accumulated phase errors are significantly suppressed compared to a single isolated mode. Interestingly, this means that in the regime of interest, i.e. $\lambda\sim \Delta_0$, the thermal Stark-shift errors in a large phononic lattice decrease to a level of $\mathcal{E}_\Lambda\lesssim 10^{-4}$, such that the use of the CCDD technique is no longer necessary.

\section{Conclusion}
In summary, we have analyzed the implementation of phonon-mediated interactions between spin qubits associated with the SiV defect center in diamond. Specifically, we have shown how the use of CDD techniques can be used to suppress the effect of ubiquitous low-frequency noise, while still permitting a flexible design of effective spin-spin interactions. A detailed analysis of all the relevant noise sources predicts that by using this approach quantum gate operations with residual errors in the range of $10^{-4}$ can be implemented in either a single mode setting or in a scalable phononic lattice scenario. 

While a quantum coherent coupling between individual spins and phonons has not been achieved in experiments yet, our results show that already with demonstrated mechanical quality factors and moderate spin-phonon interaction strengths, quantum operations with fidelities comparable to other quantum computing platforms become feasible. These high fidelities are directly related to the extremely large (protected) cooperativities $\mathcal{C}_\Omega\sim 10^5-10^7$ that can be realistically obtained for phononic modes, but are currently far out of reach for alternative coupling schemes based on optical cavities. Therefore, despite many obstacles that still need to be overcome, this unique property makes phononic quantum channels a promising approach to scale up spin-qubit based quantum computing and quantum simulation schemes in the near future.

\section{Acknowledgements}
We thank Yiwen Chu, Patrice Bertet and Marko Loncar for many stimulating discussions and acknowledge initial contributions to this project by Xue-Jian Sun during an internship, which have been published independently in Ref.~\cite{Sun22}. This work was supported by the European Union's Horizon 2020 research and innovation program under grant agreement No. 899354 (SuperQuLAN) and by Grant 62179 of the John Templeton Foundation. This research is part of the Munich Quantum Valley, which is supported by the Bavarian state government with funds from the Hightech Agenda Bayern Plus.


\appendix

\setcounter{figure}{0}
\renewcommand{\figurename}{Fig.}
\renewcommand{\thefigure}{A\arabic{figure}}

\section{Spin Qubit}\label{app:Eigenstates}
The qubit states are obtained by a numerical diagonalization of the full ground state Hamiltonian $H=H_{\rm SO}+ H_{\rm strain}+ H_{\rm Z}$. Using the basis of spin-orbit eigenstates $\{ \ket{e_-, \downarrow}, \ket{e_+, \uparrow}, \ket{e_+, \downarrow}, \ket{e_-, \uparrow}\}$, its matrix form reads~\cite{HeppThesis,Meesala2018} $(\hbar=1)$
\begin{widetext}
\begin{equation}
H=
\begin{bmatrix}
 -\lambda_{\rm SO}/2-(\gamma_S+\gamma_{\rm L})B_z& 0 & \Upsilon_x + i\Upsilon_y &  \gamma_SB_x\\
  0 &  -\lambda_{\rm SO}/2+(\gamma_S+\gamma_{\rm L} )B_z & \gamma_S B_x & \Upsilon_x - i\Upsilon_y \\
   \Upsilon_x - i\Upsilon_y  &  \gamma_SB_x & \lambda_{\rm SO}/2-(\gamma_S-\gamma_{\rm L})B_z & 0 \\
\gamma_SB_x & \Upsilon_x + i\Upsilon_y  & 0 &   \lambda_{\rm SO}/2+(\gamma_S-\gamma_{\rm L})B_z
\end{bmatrix},
\end{equation}
\end{widetext}
where $\gamma_{\rm L}=q\mu_B$ and $\gamma_S=g_e\mu_B/2$. Each of the eigenstates $\{|0\rangle,|1\rangle,|2\rangle,|3\rangle\}$ of $H$ is, in general, a combination of all spin-orbit eigenstates. Our qubit will be defined by the two lowest eigenstates $|0\rangle$ and $|1\rangle$, whose energy difference $\omega_{10}$ is on the order of a few GHz, while $\omega_{20}$ or $\omega_{30}$ are on the order of $\lambda_{\rm SO}\gg\omega_{10}$. As both the driving field (with frequency $\omega_d$) and the phonon mode (with frequency $\omega_{\rm ph}$) are far off resonant to the $|0\rangle\leftrightarrow |2\rangle$ and $|0\rangle\leftrightarrow |3\rangle$ transitions, these states can be neglected from the dynamics using the rotating wave approximation (RWA). Formally, for this RWA to be valid we require the conditions $\Omega_{n0}(t)\ll\omega_{n0}$ and $ \delta\omega_{n0}(t)\ll\omega_{n0}$ to hold for $n=2$ and $n=3$. Here, $\Omega_{n0}(t)= |\langle0|S_x|n\rangle|g_e\mu_B  \mathcal{B}(t)/\hbar$ and $\delta\omega_{n0}(t) =   |\langle0|S_z|n\rangle| g_e \mu_B  \mathcal{B}(t)/\hbar$. For typical parameters assumed in this paper, $\Omega_{n0}(t)/(2\pi) \sim \delta\omega_{n0}(t)/(2\pi) \lesssim 50$ MHz while $\omega_{n0}/(2\pi)\sim 50 $ GHz.  Thus, these conditions are fully satisfied.

\section{Effect of longitudinal and counter-rotating terms}\label{app:LongitudinalCouplings}
\begin{figure}
	\includegraphics[width=\columnwidth]{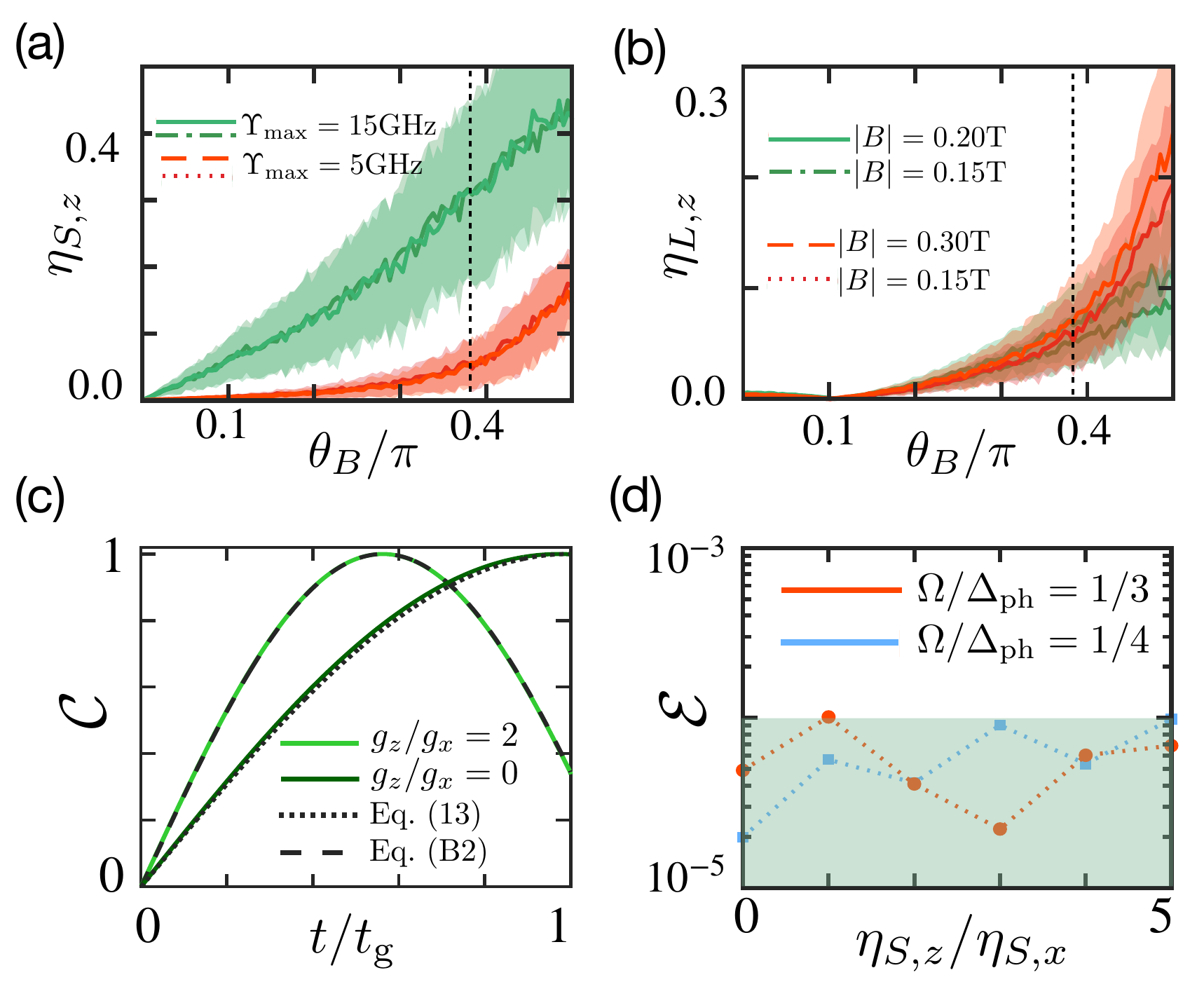} 
	\caption{ Influence of longitudinal terms in two-qubit gates. The plots in (a) and (b) are analogous to the plots in Figs.~\ref{fig2}~(b) and (c) respectively, but for the longitudinal matrix elements $\eta_{S,z}$ and $\eta_{L,z}$. (c) Concurrence versus time, calculated using $H_{\rm full}$ for values of $g_z/g_x=0$ (dark solid line) and $g_z/g_x=2$ (bright solid line). In addition, we plot the concurrence according to Hamiltonian~\eqref{eq:HQubitPhonon}, and the corrected effective model in Eq.~(\ref{Eq:H_effective_corrected}) (for $g_z/g_x=2$). Note that the longitudinal coupling term going with $\eta_{L,z}$ can significantly affect the instant in which a maximally entangled state is achieved. (d) Infidelity versus the value of the longitudinal matrix element $\eta_{S,z}$, after evolving $H_{\rm full}$ for a time $t_{\rm g}=\pi/4\tilde J_\perp$, with $g_z/g_x=2$ and $\phi_S=0$, and for the gate parameters used in Sec.~\ref{subsec:Optimization}, with a fixed ramp duration $t_{\rm ramp}=0.5\times 2\pi/\Delta_{\rm ph}$.  }\label{figS1}
\end{figure}

When approximating Eq.~\eqref{Eq:HSiVTLS} with the final qubit Hamiltonian in Eq.~\eqref{Eq:drivenSiV}, but also when deriving the Jaynes-Cummings-type spin-phonon coupling given in  Eq.~\eqref{Eq:JCHamiltonian}, we have made a RWA by neglecting the counter-rotating terms. In general, this approximation is valid for $\Omega, g \ll  \omega_{10},\omega_d$, but for the considered parameters, these terms might still induce relevant corrections to the gate fidelities at the level of $10^{-4}$. In addition, in both steps of the derivation we have omitted longitudinal interactions $\sim\eta_{S,z}, \eta_{L,z}$. In Fig.~\ref{figS1} (a) and (b) we plot those matrix elements under the same conditions as assumed in Fig.~\ref{fig2} in the main text. We see that for the same values of strain and magnetic fields, the longitudinal strain coupling matrix element $\eta_{L,z}$ achieves values similar to $\eta_{L,x} \sim 0.1$, while the longitudinal spin transition matrix element $\eta_{S,z}$ can reach values of up to about five times $\eta_{S,x}$. Therefore, also these contributions give rise to off-resonant interactions that must be taken into account.

To evaluate the potential corrections from all those effects on the predicted gate fidelities, we consider  the full Hamiltonian $H_{\rm full}= \omega_{\rm ph}a^\dagger a +\sum_{i=1,2} H^{(i)}_{\rm SiV}+ H^{(i)}_{\rm e-ph}$, where $H_{\rm SiV}^{(i)}$ is the qubit Hamiltonian without RWA given in Eq.~\eqref{Eq:HSiVTLS} and 
\begin{equation}
H_{\rm e-ph}^{(i)}=\left(g_x\sigma_i^x+g_z\sigma_i^z\right)(a+a^\dagger).
\end{equation}
Here, $g_{x}=  d \epsilon_{\rm ph} \eta_{L,x}$ and $g_{z}= d \epsilon_{\rm ph} \eta_{L,z}$, assuming identical parameters for both qubits and $\eta_{L,x}\in \mathbbm{R}$ for simplicity. 

Based on this full Hamiltonian, we re-simulate the implementation of a phonon-mediated two qubit gate under similar conditions as in Sec.~\ref{sec:DressedQubits}. In Fig.~\ref{figS1}~(c) we show, first off all, the evolution of the concurrence $\mathcal{C}$ for different ratios of $\eta_{L,z}/\eta_{L,x}$, but assuming $\eta_{S,z}=0$. Already for $\eta_{L,z}=0$ we see small differences between the predictions of the full model (solid dark line) and Hamiltonian~\eqref{eq:HQubitPhonon} (dotted line), which is based on a RWA. Although these deviations are barely visible on the scale of this plot, they clearly lead to deviations far above the targeted level of $10^{-4}$. More strikingly, for the case $\eta_{L,z}/\eta_{L,x}=2$, the time to produce a Bell state ($\mathcal{C}=1$) is almost a factor of two shorter than predicted by Eq.~\eqref{eq:HQubitPhonon}. 

These simulations show that in the regime of interest, the off-resonant contributions we neglected in the main text result in significant corrections to the two-qubit dynamics. Fortunately, these deviations are not essential and can be reabsorbed into a renormalization of the parameters of the effective qubit Hamiltonian using second-order perturbation theory. After a lengthy calculation we find that in an interaction picture with respect to $H_0=\sum_i\omega_{10}{\sigma}_i^z/2$ this corrected effective Hamiltonian is equivalent to Eq.~\eqref{Eq:H_effective},
\begin{equation}\label{Eq:H_effective_corrected}
\begin{split}
H_{I,{\rm eff}}=&\sum_i   \frac{\tilde \Omega}{2}\tilde{\sigma}_i^z-\tilde{\Lambda}(a^\dagger a+1/2)\tilde\sigma_i^z\\
&-\tilde{J}_{\perp}(\tilde\sigma_1^+\tilde\sigma_2^-+\tilde\sigma_1^-\tilde\sigma_2^+) - \frac{\tilde{J}_{\parallel}}{2}\tilde\sigma^z_1\tilde\sigma_2^z,
\end{split}
\end{equation}
but with a reduced Rabi frequency 
\begin{equation}
\tilde \Omega =\Omega \left[1-\frac{\Omega^2/4+\delta\omega^2}{8\omega_{10}^2}\right]
\end{equation}
and modified couplings
\begin{equation}\label{Eq:spinspincouplings}
\tilde{J}_{\perp}=\sum_{\nu=1,2,3} \Big\{ \frac{ |g^{-}_\nu|^2}{\Delta_{\nu}-\Omega}+\frac{|g^{+}_\nu|^2}{\Delta_{\nu}+\Omega} \Big\}, \qquad \tilde{J}_\parallel=\sum_{\nu=1,2,3} \frac{|g_\nu^\parallel|^2}{\Delta_{\nu}}
\end{equation}
and
\begin{equation}
\tilde{\Lambda}=\sum_{\nu=1,2,3} \Big\{ \frac{ |g^{-}_\nu|^2}{\Delta_{\nu}-\Omega}-\frac{|g^{+}_\nu|^2}{\Delta_{\nu}+\Omega} \Big\}.
\end{equation} 
Here, $\Delta_1=\Delta_{\rm ph}$, $\Delta_2=2\omega_{\rm ph}-\Delta_{\rm ph}$ and $\Delta_3=\omega_{\rm ph}$, $g^{\pm}_{1}=\frac{1}{2}g_x\chi_0(\tilde{\chi}_0 \pm \tilde{\chi}_1e^{i2\phi_S})$, $g^{\pm}_{2}=\frac{1}{2}g_x\chi_0\tilde{\chi}_0$, and $g^{\pm}_{3}=g_z\tilde{\chi}_0$, and $g^{\parallel}_{1,2}=g_x\chi_0$ and $g^{\parallel}_{3}=0$. Also, $\chi_0=1-(\delta\omega/\omega_{10})^2$, $\tilde{\chi}_0=1-\Omega^2/(4\omega_{10}^2)$, and $\tilde{\chi}_1=\Omega/(4\omega_{10})$. 

In Fig.~\ref{figS1}~(c), the dashed line shows the evolution of the concurrence as a function of time for Hamiltonian~\eqref{Eq:H_effective_corrected}, for $\eta_{L,z}/\eta_{L,x}=2$. Note that this matches the evolution given by $H_{\rm full}$ and thus validates the effective Hamiltonian in Eq.~\eqref{Eq:H_effective_corrected}. To further check the equivalence, we define our gate according to Hamiltonian~\eqref{Eq:H_effective_corrected}, with gate time $t_{\rm g}=\pi/(4\tilde J_\perp)$, and compare its evolution with that of $H_{\rm full}$ by using the state fidelity. In particular, in Fig.~\ref{figS1}~(d), we simulate the full Hamiltonian to calculate the resulting error $\mathcal{E}$ as a function of $\eta_{S,z}/\eta_{S,x}$, and for two different values of the Rabi frequency $\Omega=\Delta_{\rm ph}/4$ and $\Omega=\Delta_{\rm ph}/3$, with a fixed ramp time $t_{\rm ramp}=0.5\times2\pi/\Delta_{\rm ph}$. These parameters are close to the optimal values identified in Fig.~\ref{Fig:EntError} (a). In addition, we choose $\eta_{L,z}/\eta_{L,x}=2$ and set $\phi_{S}=0$ for concreteness. We see that for these parameters the full model agrees with the predictions of the effective Hamiltonian up to errors of $\mathcal{E}\leq 10^{-4}$ for all values $\eta_{S,z}/\eta_{S,x}\leq 5$, without any other adjustments. 

We conclude that while making a RWA and neglecting off-resonant coupling terms is not strictly valid in our setup, the corrections are purely coherent and can be accounted for by a renormalization of the effective model parameters. Since these adjustments do not significantly affect any of the incoherent errors, they do not change any of the conclusion derived in the main part of the paper.

\section{Derivation of the effective spin models}\label{app:SWTrans}

A common technique to derive phonon-mediated effective spin-spin interactions is based on the Schrieffer-Wolff  transformation (SWT). For any Hamiltonian of the form $H=H_0+V$, where $H_0$ ($V$) is the free (interaction) term, the SWT implements a change of basis states via  the unitary transformation $e^{iS}$, where $S$ fulfills the condition $[S,H_0]=iV$. If the interaction is small, i.e., $||V||\ll||H_0||$, the transformed Hamiltonian takes the form $H_{\rm eff}= e^{iS}He^{-iS}\approx H_0+\frac{i}{2}[S,V]$, which is valid up to second order in $V$.

In our case, the Hamiltonian describing the interaction of two spins with the cavity mode is given by 
\begin{equation}
H_0+V= \sum_{i=1}^N \frac{\Omega_i}{2}\tilde{\sigma}_i^z+\Delta_{\rm ph} a^\dagger a+ g_i \left[ a^\dag  \Sigma_i  + a \Sigma_i^\dag \right].
\end{equation}
In addition, to account for the effect of dissipation, we include the coupling of the phonon mode to a Markovian environment, 
\begin{equation}\label{Eq:cavbathHamil}
H_{c-b}(t)= 
\sqrt{\kappa}\left[ae^{-i\omega_{10} t}B^\dagger(t) + a^\dagger e^{i\omega_{10} t}B(t)\right],
\end{equation}
where the bosonic operators $B(t)$ and $B^\dag(t)$, with $[B(t),B^\dag(t')]\sim \delta(t-t')$,  represent short-lived excitations in the bath. 

After applying the SWT to the full Hamiltonian, the resulting transformed Hamiltonian  
$\tilde H_{\rm eff}=e^{iS}[H_0+V+H_{c-b}(t)]e^{-iS}$ reads
\begin{equation}\label{PreEffEq}
\tilde H_{\rm eff}\approx H_{\rm eff}+H_{c-b}(t) +\sqrt{\kappa}(i[S,a]B^\dagger(t) e^{-i\omega_{10}t}+ {\rm H.c.}),
\end{equation}
where the first term contains the  effective spin-spin interaction 
\begin{equation}
H_{\rm eff}= H_0-\frac{1}{2} \sum_{i,j} g_ig_j \sum_{\mu,\mu'}  \zeta_\mu \gamma_{\mu'}  [ \tilde\sigma_i^\mu a , \tilde\sigma_j^{\mu'} a +{\rm H.c.}]+{\rm H.c.}, 
\end{equation}
while the last term describes the effective interaction between the spins and the bath operators $B$. Here, $S$ takes the form
\begin{equation}
S=i\sum_i \sum_{\mu=+,-,z}g_i\zeta_\mu(a\tilde\sigma_i^\mu-{\rm H.c.}),
\end{equation}
where 
\begin{equation}\label{ZetaDef}
\zeta_\mu=\frac{\gamma_\mu}{\Delta+\beta_\mu\Omega},
\end{equation}
with ${\gamma}_{+,-,z}=\cos^2{(\theta/2)},-\sin^2{(\theta/2)},\sin{(\theta)}/2$ and $\beta_{+,-,z}=-1,1,0$. 

Under the usual Born-Markov approximation, tracing out the bath leads to an effective description via the master equation 
\begin{eqnarray}
\dot{\rho}=-i[H_{\rm eff},\rho] + \sum_\mu\Gamma^-_\mu\mathcal{D}[S_\mu^\dagger]\rho+\Gamma^+_\mu\mathcal{D}[S_\mu]\rho  \nonumber\\  +\kappa(n_{\rm th}+1)\mathcal{D}[a]\rho+\kappa n_{\rm th}\mathcal{D}[a^\dagger]\rho.
\end{eqnarray}
Here, $\tilde{S}_\mu=\sum_j(g_j/g)\tilde\sigma_j^\mu$, $\Gamma^-_\mu= \kappa g^2(n_{\rm th}+1)\zeta^2_\mu$ and $\Gamma^+_{\mu}= \kappa g^2 n_{\rm th}\zeta^2_\mu$. After some simplifications, the dissipative spin part reduces to Eq.~\eqref{eq:PID} in the main text.

Finally, we note that a similar derivation can be pursued for the multimode case studied in Sec.~\ref{sec:scalability}. In that case, the original Hamiltonian reads
\begin{equation}
\begin{split}
H_{0}+V\simeq&\sum_{k=1}^{N} \Delta_kc^\dagger_kc_k+ \sum_{i=1}^{N} \frac{\tilde{\Omega}_i}{2}\tilde{\sigma}_i^z \\ &+\sum_{i,m}^Ng_{i} b_{ik} \left(c_k^\dag \Sigma^-_i +  c_k \Sigma_i^+\right),
\end{split}
\end{equation}
and the corresponding operator for the SWT is
\begin{equation}
S=i\sum_{i,k} \sum_{\mu}g_ib_{ik}\zeta_\mu(c_k\tilde\sigma_{i}^{\mu}-{\rm H.c.}),
\end{equation}
where 
\begin{equation}
\zeta_{\mu,k}=\frac{\gamma_\mu}{\Delta_k+\beta_\mu\Omega}.
\end{equation}
After a RWA, for which $J^{\perp,\parallel}_{ij},\Lambda_{jk}\ll |\Delta_k-\Delta_{k'}|,|\Delta_k|, |\Delta_k\pm\Omega_i|$ is required, we obtain the effective spin Hamiltonian in Eq.~\eqref{Eq:SpinSpinHam}. The master equation then reads 
\begin{eqnarray}
\dot{\rho}=-i[H_{\rm eff},\rho] + \sum_{\mu,k}\Gamma^-_{\mu,k}\mathcal{D}[S_{\mu,k}^\dagger]\rho+\Gamma^+_{\mu,k}\mathcal{D}[S_{\mu,k}]\rho  \nonumber\\  +\sum_{k}\kappa(n_{\rm th}+1)\mathcal{D}[c_k]\rho+\kappa n_{\rm th}\mathcal{D}[c_k^\dagger]\rho, 
\end{eqnarray}
where, $\tilde{S}_{\mu,k}=\sum_i(g_i/g)b_{ik}\tilde\sigma_i^\mu$, $\Gamma^-_{\mu,k}= \kappa g^2(n_{\rm th}+1)\zeta^2_{\mu,k}$ and $\Gamma^+_{\mu,k}= \kappa g^2 n_{\rm th}\zeta^2_{\mu,k}$. Note that here we considered all normal modes to be in an stationary thermal state with an average number of phonons $n_{\rm th}$, which is a good approximation if $4\lambda \ll \omega_{\rm ph}$.

\section{Error estimates}\label{app:analyticerrors}
As the basis to calculate the influence of the different errors on the spin dynamics, we assume that the latter can be described by the Hamiltonian $H=H_0+H_{\rm g}+ H_{E}(t)$. Here $H_0=\tilde \Omega \sum_i \tilde \sigma_i^z /2 +  \Delta_{\rm ph} a^\dag a$, $H_{\rm g}$ is the ideal gate Hamiltonian given in Eq.~\eqref{Eq:gateHamil}, and 
$H_E(t)$ accounts for the additional perturbation that we want to investigate. By assuming that the influence of $H_E(t)$ on the spin dynamics is small, we can write the qubit state at the end of the gate sequence as $\rho_q(t_{\rm g})\approx\rho_{\rm id} +\Delta\rho_q(t_{\rm g})$, where $\rho_{\rm id}=e^{i (H_0+H_{\rm g})t_{\rm g}}\rho_q(0)e^{-i (H_0+H_{\rm g})t_{\rm g}}$ is the targeted state. The deviation $\Delta\rho_q(t_{\rm g})$ can be estimated in second-order perturbation theory by the Nakajima-Zwanzig equation 
\begin{widetext}
\begin{equation}\label{PerturbativeNZ}
\Delta\rho_q (t_{\rm g})\approx-\int_0^{t_{\rm g}}dt'\int_0^{t'}dt'' {\rm Tr}_{\rm ph}\{[H'_E(t'),[H'_E(t''),\rho_q(0)\otimes \rho_{\rm th}]]\} .
\end{equation}
\end{widetext}
where $H'_{E}(t)=e^{i (H_0+H_{\rm g})t}H_{E}(t)e^{-i (H_0+H_{\rm g})t}$. The corresponding error contribution is then given by 
\begin{equation}
\mathcal{E}(t_{\rm g})=|{\rm Tr}_q\{\rho_q(0) \Delta\rho_q (t_{\rm g}) \}|.
\end{equation}

In the main text, we discuss three types of error whose influence we can calculate using Eq.~\eqref{PerturbativeNZ}: (i) phonon-induced decoherence, characterized in Appendix~\ref{app:SWTrans}, (ii) the influence of magnetic field fluctuations, discussed in Secs.~\ref{sub:lowfreq} and~\ref{sub:dressedqubits}, and (iii) the influence of the thermal Stark shift. In order to carry out these calculations explicitly, we make another simplification by replacing $H_E^\prime(t)\rightarrow e^{i H_0 t}H_{E}(t)e^{-i H_0 t}$. This simplifies the calculations, but ignores the influence of the gate Hamiltonian on the qubit evolution and the error is only estimated by its action on the initial state. Therefore, to compensate for this approximation, we evaluate the error for both the initial state $\rho_q(0)$ and the final state $\rho_{\rm id}$, and take the average.

\subsection{Phonon-induced spin decoherence}
In the case of the phonon-induce spin decoherence, we identify $H_E(t)$ with the effective spin-bath interaction described by the last term in Eq.~\eqref{PreEffEq}, and 
Eq.~\eqref{PerturbativeNZ} simplifies to 
\begin{equation}
\Delta\rho_q (t_{\rm g})\approx t_{\rm g}\sum_{\mu=\pm,z}\sum_{i,j}\Gamma_{ij}^{\mu} \mathcal{D}_{i,j}[\tilde{\sigma}_\mu] \rho_q(0).
\end{equation}
For $g_i=g_j=g$ and a pure initial state, the error reads
\begin{equation}
\mathcal{E}_\kappa\approx t_{\rm g} \left|\sum_\mu \Gamma_\mu \sum_{ij}  \langle\tilde\sigma_i^\mu \rangle \langle(\tilde\sigma_j^\mu)^\dagger\rangle -\langle(\tilde\sigma_j^\mu)^\dagger\tilde\sigma_i^\mu\rangle\right|.
\end{equation}
For both the initial and the final state, $|\tilde 0 \tilde 1\rangle $ and $|\Psi\rangle$, the result is $\mathcal{E}_\kappa\approx (\Gamma_++ \Gamma_-)t_{\rm g}$, where 
\begin{equation}
\Gamma_++ \Gamma_-=\frac{ g^2\kappa(2n_{\rm th}+1)}{4} \left[ \frac{1}{(\Delta_{\rm ph}-\Omega)^{2}} +\frac{1}{(\Delta_{\rm ph}+\Omega)^{2}} \right]
\end{equation}

\subsection{Magnetic noise}

In the case of external magnetic field fluctuations, we assume the fluctuating field $\xi_i(t)$ to be static during the duration of the gate, i.e., $H_{E}(t)=\sum_j B_j\tilde{\sigma}_j^z$, where $B_j=\delta\xi_j^2/(4\Omega)$, with $\delta\xi_i^2=\xi_i^2-\langle\xi_j^2\rangle$.   Eq.~\eqref{PerturbativeNZ} then simplifies to
 \begin{eqnarray}
\Delta\rho_q (t_{\rm g})\approx - t_{\rm g}^2 \sum_{i,j} \langle B_iB_j \rangle  \mathcal{D}_{i,j}[\tilde{\sigma}_z] \rho_q(0),
\end{eqnarray}
where $\langle B_iB_j \rangle = \frac{\delta_{ij}}{(4\Omega)^2}(\langle\xi_j^4\rangle-\langle\xi_j^2 \rangle^2)$. For Gaussian variables $\langle\xi_j^4\rangle=3\sigma^4$ and $\langle\xi_j^2\rangle=\sigma^2$. For the state $|\Psi\rangle$, the resulting error is $\mathcal{E}_\varphi\approx t_{\rm g}^2\sigma^4/(4\Omega^2)$, while for the state $|\tilde 0\tilde 1\rangle$ the result is zero. The mean value thus gives $\mathcal{E}_\varphi\approx t_{\rm g}^2\sigma^4/(8\Omega^2)=\frac{1}{2}(t_{\rm g}/T_2^\Omega)^2$. 

\subsection{Stark-shift dephasing}\label{subapp:StarkDeph}
In the case of the Stark-shift dephasing, Eq.~\eqref{PerturbativeNZ} simplifies to
 \begin{eqnarray}
\Delta\rho_q (t_{\rm g})&\approx&- \sum_{ij}\Lambda_i\Lambda_j \mathcal{D}_{i,j}[\tilde{\sigma}_z] \rho_q(0) \nonumber \\
&\times&\int_0^{t_{\rm g}}dt'\int_0^{t'}dt''  \langle \delta \hat{n}(t'') \delta \hat{n}(0)\rangle,
\end{eqnarray}
where we used the stationarity of the bosonic state, i.e. $ \langle \delta n(t') \delta n(t'')\rangle = \langle \delta n(t'') \delta n(t')\rangle = \langle \delta n(t'-t'') \delta n(0)\rangle $. For a thermal oscillator
\begin{equation}\label{Eq:correlationSinglemode}
\langle  \delta n(\tau)\delta n(0)\rangle= n_{\rm th}(n_{\rm th}+1)e^{-\kappa \tau}, 
\end{equation}
and  
\begin{equation}
\int_0^{t_{\rm g}} dt'\int_0^{t'} d\tau\langle  \delta n(\tau)\delta n(0)\rangle= \frac{n_{\rm th}(n_{\rm th}+1)}{\kappa^2}[e^{-\kappa t_{\rm g}}-(1-\kappa t_{\rm g})]. 
\end{equation}
For $\Lambda_i=\Lambda_j=\Lambda$ and for the state $|\Phi\rangle$, we then obtain the result in Eq.~\eqref{Eq:StarkError}.

\section{Residual spin-phonon entanglement}\label{app:SPEerror}
The Hamiltonian that governs the two-qubit interaction is of the form
\begin{equation}
H=\Delta_{\rm ph}a^\dagger a+ \sum_{i=1}^N \frac{\Omega_i}{2}\tilde{\sigma}_i^z+ g_i \left[ a^\dag  \Sigma_i  + a \Sigma_i^\dag \right],
\end{equation}
which, after the SWT, becomes $H_{\rm eff}$. While the actual evolution followed by the system is given by $\rho(t)=e^{-i H t}\rho_{\rm ini} e^{i H t}$, in our approximation we assume that the state evolves as $\rho_{\rm eff}(t)=e^{-i H_{\rm eff} t}\rho_{\rm ini}  e^{i H_{\rm eff} t}$ instead. The error associated with this inaccuracy can be quantified by calculating the state infidelity between $\rho(t)$ and $\rho_{\rm eff}(t)$, i.e. $\mathcal{E}_{\rm s-ph}=1-|{\rm Tr}\{\rho_{\rm eff} (t) \rho(t)\}|$. Since $H_{\rm eff}=e^{i S}He^{-iS}$, this gives 
\begin{equation}
\mathcal{E}_{\rm s-ph}=1-\left|{\rm Tr}\{\rho_{\rm ini} e^{-iS_I} e^{iS} \rho_{\rm ini} e^{-iS} e^{iS_I}\}\right|,
\end{equation}
where $S_I(t)=e^{iH_{\rm eff}t} S e^{-iH_{\rm eff}t}$. Note that, if $S_I(t_{\rm g})=S$, the error associated with the residual spin-phonon entanglement vanishes, $\mathcal{E}_{\rm s-ph}=0$. As a worst-case scenario, we take $S_I(t_{\rm g})$ to be $-S$. Then, 
\begin{equation}
\mathcal{E}_{\rm s-ph}\approx1-\left|{\rm Tr}\{\rho_{\rm ini}  e^{i2S} \rho_{\rm ini} e^{-i2S}\}\right|.
\end{equation}
Expanding the exponentials in powers of $S$ and using ${\rm Tr}_{\rm ph}\{(a^\dagger)^n\rho_{\rm th}\}=0={\rm Tr}_{\rm ph}\{a^n\rho_{\rm th}\}$ and ${\rm Tr}_{\rm ph}\{a^\dagger a\rho_{\rm th}\}=n_{\rm th}$, we finally obtain
\begin{eqnarray}
\mathcal{E}_{\rm s-ph}\approx 4g^2\Big| \sum_{\mu,\mu'}\zeta_\mu\zeta_{\mu'}n_{\rm th}\Big[\langle \tilde S_\mu\rangle\langle \tilde S_{\mu'}^\dagger\rangle- \langle \tilde S_{\mu'}^\dagger \tilde S_\mu\rangle \Big] \\ +\zeta_\mu\zeta_{\mu'}  (n_{\rm th}+1)\Big[\langle \tilde S_\mu^\dagger\rangle\langle \tilde S_{\mu'}\rangle -\langle \tilde S_{\mu'} \tilde S_\mu^\dagger\rangle \Big]\Big|,
\end{eqnarray}
where $\tilde{S}_\mu$ and $\zeta_\mu$ follow the definition given in Appendix~\ref{app:SWTrans}. For the spin state $|\tilde 0\tilde 1\rangle$, this gives
\begin{equation}
\mathcal{E}_{\rm s-ph}\approx4g^2(\zeta_+^2+\zeta_-^2)(2n_{\rm th}+1),
\end{equation}
and, for $\theta=\pi/2$, 
\begin{equation}
\mathcal{E}_{\rm s-ph}\approx g^2(2n_{\rm th}+1)\left[\frac{1}{(\Delta_{\rm ph}-\Omega)^2}+\frac{1}{(\Delta_{\rm ph}+\Omega)^2}\right].
\end{equation}
With the help of the adiabatic ramp for $\Omega(t)$, the error can be reduced to $\mathcal{E}_{\rm s-ph}\approx (g/\Delta_{\rm ph})^2(2n_{\rm th}+1)$.


\section{Parameter optimization}\label{app:Minimization}
In this appendix we summarize the derivation of the optimal driving parameters and the resulting minimal gate error. We first do so for a conventional, i.e, undriven, cavity QED system with Hamiltonian $H=\Delta_{\rm ph} a^\dagger a  + \sum_j g(\sigma_j^+a+\sigma_j^-a^\dagger)$. In this case we obtain the effective interaction $H_{\rm eff}=-J_\perp(\sigma_1^+\sigma_2^-+\sigma_1^-\sigma_2^+)$ with $J_\perp=g^2/\Delta_{\rm ph}$. At time $t_{\rm g}=\pi/4 J_\perp$, $H_{\rm eff}$ transforms the state $|e, g\rangle$ into the entangled state $|\Psi\rangle=(|e,g\rangle+i|g,e\rangle)/\sqrt{2}$. For static dephasing, an error estimate similar to the one above gives 
\begin{equation}
\mathcal{E}\simeq  \frac{g^2\kappa}{\Delta_{\rm ph}^2} (2n_{\rm th}+1)  t_{\rm g} + \frac{1}{2} (t_{\rm g}/T_2^*)^2.
\end{equation}
A minimization of this expression with respect to the detuning $\Delta_{\rm ph}$ yields an optimal value of $\Delta_{\rm opt}=(4\kappa(2 n_{\rm th}+1)g^4(T_2^*)^2/\pi)^{1/3}$. After inserting this value back into the expression for $\mathcal{E}$ we obtain the result stated in Eq.~\eqref{eq:minErrorUndriven}.

Let us now consider the case of dressed qubits, as discussed in Sec.~\ref{subsec:Optimization}.
For a fixed detuning $\Delta_{\rm ph}$, we aim at minimizing the error term $\mathcal{E}_\varphi+\mathcal{E}_\kappa$, which can be written as
\begin{equation}\label{Eq:exactError}
\mathcal{E}_{\kappa+\varphi}=\mathcal{E}_\kappa^{\min}\frac{1+x^2}{1-x^2} + \mathcal{E}_\varphi^{ 0} \frac{(1-x^2)^{2}}{x^2},
\end{equation}
where $\mathcal{E}_\kappa^{\min}=\pi\kappa(2n_{\rm th}+1)/(4\Delta_{\rm ph})$, $\mathcal{E}_\varphi^{ 0}=\frac{\pi^2}{8}(gT_2^*)^{-4}$, and $x=\Omega/\Delta_{\rm ph}$. The optimal value of $x_{\rm opt}$, for which $\mathcal{E}_{\kappa+\varphi}$ has its minimum, can be found exactly. However, the corresponding expression is too complicated to be of use. Therefore, we focus on the limit $x\approx 1$, where $(\Delta_{\rm ph}-\Omega) \ll \Delta_{\rm ph}$. Physically, this limit is most relevant when both the detuning $\Delta_{\rm ph}$ and the Rabi frequency $\Omega$ can be made arbitrarily large, and, thus, $\mathcal{E}_\kappa^{\min}\ll\mathcal{E}_\varphi^{ 0} $. In this limit, we rewrite $x$ as $x=(1+r)^{-1}$ with $r\ll1$, and we obtain 
\begin{equation}\label{Eq:exactError2}
\mathcal{E}_{\kappa+\varphi}\approx\frac{\mathcal{E}_\kappa^{\min}}{r}+ 4\mathcal{E}_\varphi^{0} r^2,
\end{equation}
whose minimum,
\begin{equation}
\mathcal{E}_{\kappa+\varphi}^{\rm min}\approx 3({\mathcal{E}_\kappa^{\min}}^2\mathcal{E}_\varphi^{0})^{1/3},
 \end{equation} 
is obtained at $r_{\rm opt}=\frac{1}{2}(\mathcal{E}_\kappa^{\min}/\mathcal{E}_\varphi^{0})^{1/3}$. This is consistent with our assumption $\Omega\approx \Delta_{\rm ph}$, for which we then obtain the result in Eq.~\eqref{eq:Error_Min}.

\begin{figure}
	\includegraphics[width=0.75\columnwidth]{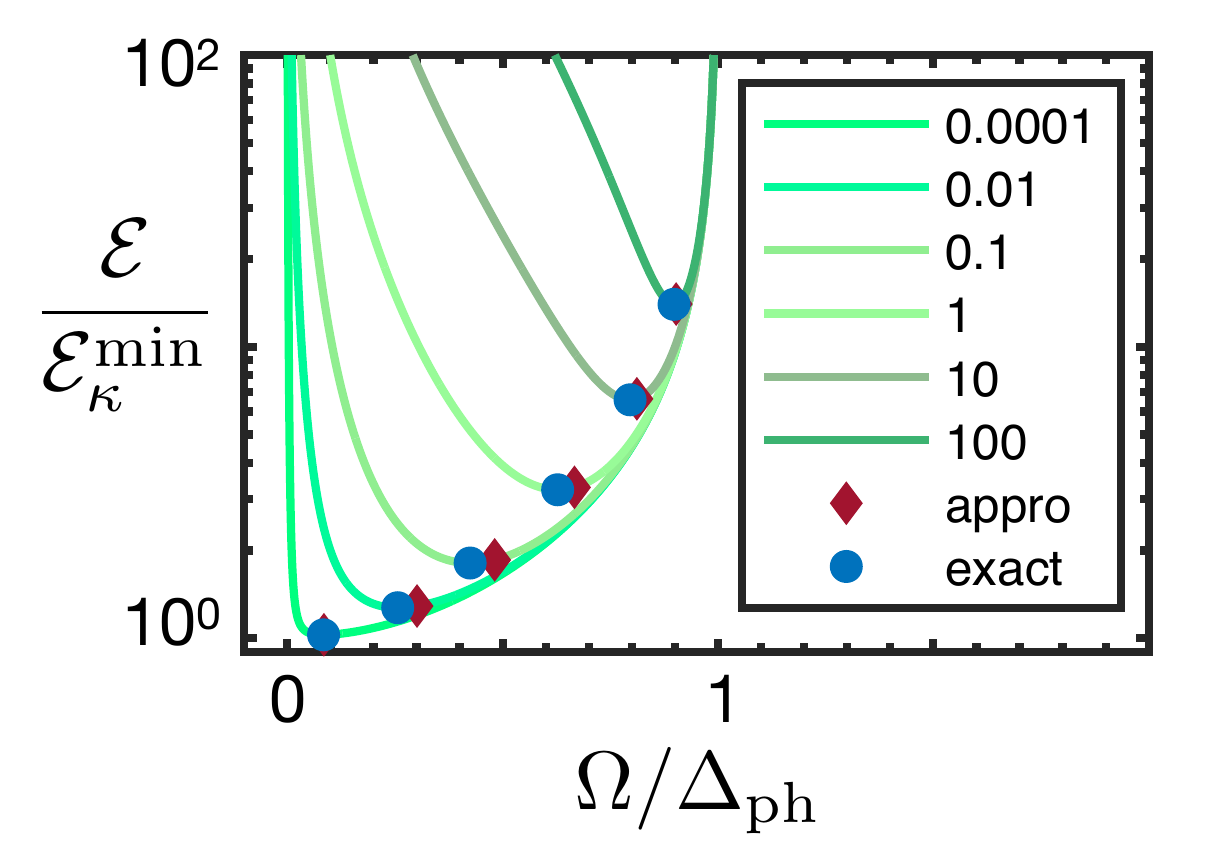} 
	\caption{ The green lines represent $\mathcal{E}_{\kappa+\varphi}$ (normalized by $\mathcal{E}_\kappa^{\min}$) versus $x=\Omega/\Delta_{\rm ph}$ for different values of $\mathcal{E}_\kappa^{\min}/\mathcal{E}_\varphi^{ 0}=[10^{-4},10^{-2},10^{-1},10^{0},10^{1},10^{2}]$. Round markers represent the exact minima, while diamond markers represent the approximate minima at $\Omega_{\rm opt}/\Delta_{\rm ph}=(1+r_{\rm opt})^{-1}$ given in Eq.~(\ref{eq:Error_Min}). }\label{figS3}
\end{figure}

In Fig.~\ref{figS3}, we plot $\mathcal{E}_{\kappa+\varphi}$ (normalized by $\mathcal{E}_\kappa^{\min}$) versus the ratio $\Omega/\Delta_{\rm ph}$ (solid lines) for different ratios of $\mathcal{E}_\kappa^{\min}/\mathcal{E}_\varphi^{0}$. The minimal error obtained for each curve (blue dots) is compared with the approximate analytic prediction from above (red diamonds). Although this result has been derived under the assumption $\mathcal{E}_\kappa^{\min}/\mathcal{E}_\varphi^{ 0} \ll1$, it gives a very accurate prediction over a wide range of parameters. Thus we conclude that Eq.~\eqref{eq:Error_Min} provides a very reliable estimate for the minimally achievable gate error, given that there are no constraints on $\Omega$ and $\Delta_{\rm ph}$ and that the spin-phonon entanglement error, $\mathcal{E}_{\rm s-ph}$, is negligible.

\section{Adiabatic ramps}\label{app:adiabaticramps}
\begin{figure}
	\includegraphics[width=\columnwidth]{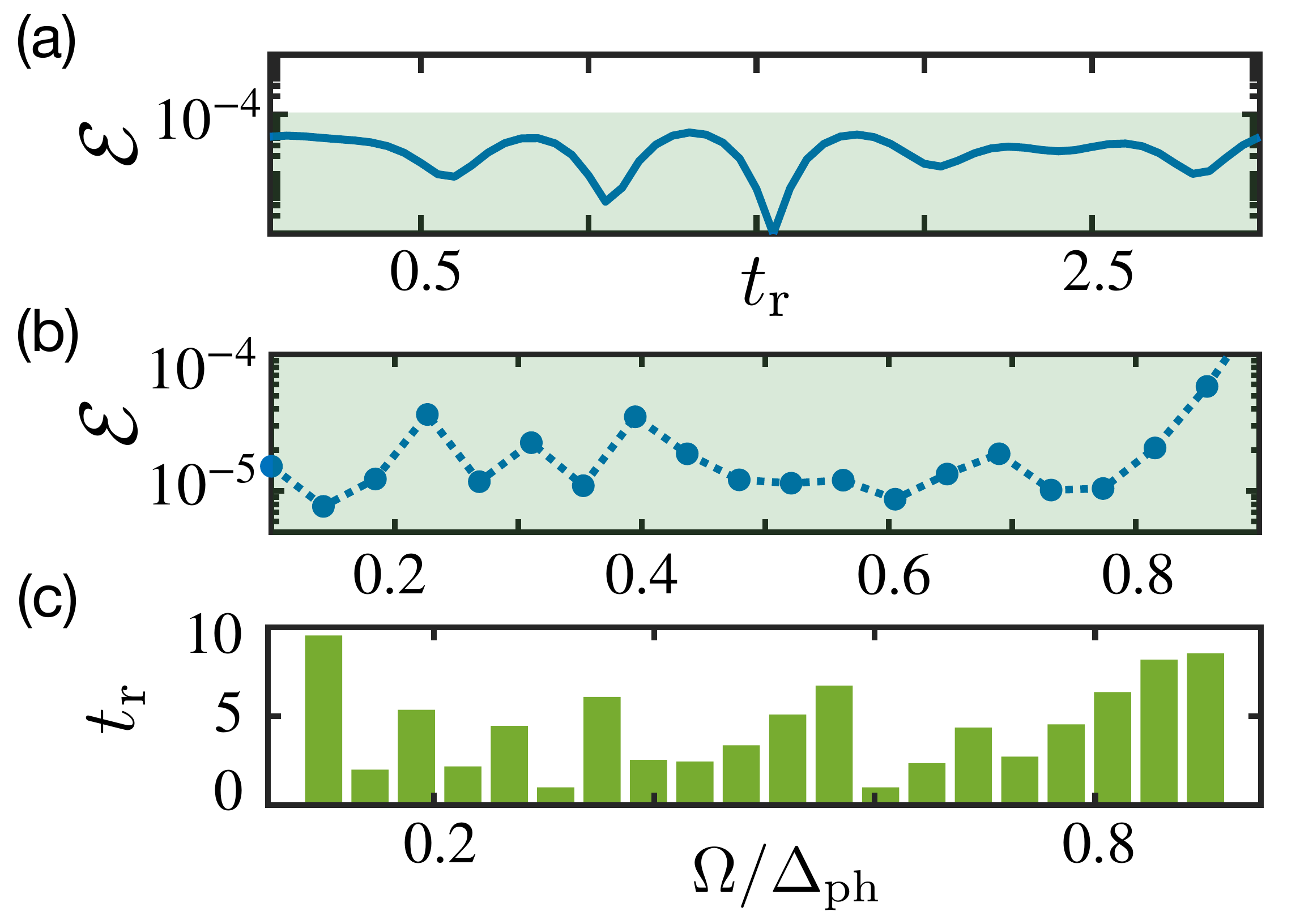} 
	\caption{Adiabatic ramps. (a) State infidelity versus the normalized ramp time $t_{\rm r}= t_{\rm ramp} \Delta_{\rm ph}/(2\pi)$. (b) Optimized infidelity $\mathcal{E}_{\rm s-ph}^{\rm min}$ versus the Rabi frequency $\Omega$. (c) Optimized ramp time  $t_{\rm r}^{\rm opt}$ versus the Rabi frequency. }\label{figS4}
\end{figure}

In all our numerical simulations, we optimize the gate performance by introducing adiabatic ramps of the Rabi frequency. These ramps are assumed to be of the shape $\Omega(t)=\Omega\sin^2[\pi (t-t_m)/(2t_{\rm ramp})]$ at the beginning ($t_m=0$) and at the end ($t_m=t_{\rm g}$) of each operation. In particular, in Sec.~\ref{subsec:Entanglement}, the ramp duration $t_{\rm ramp}$ is chosen to minimize the spin-phonon entanglement error. This optimization is done by solving the Schr\"odinger equation for the time-evolution operator, $i\hbar \dot{U}(t)= H(t,t_{\rm ramp}) U(t)$, for Hamiltonian~\eqref{Eq:H_full_1mode_Dressed_IP} (without external dephasing terms), and for different values of $t_{\rm ramp}$ and $\Omega$. After calculating the reduced density matrix $\rho_q$ resulting from the application of $U(t_{\rm g})$ on $|\Psi\rangle\otimes |0\rangle_{\rm ph}$, 
we compare this state with the ideal final state $U_{\rm g}|\Psi\rangle$ using the state fidelity $\mathcal{F}$. In Fig.~\ref{figS4} (a) we show the resulting error, $\mathcal{E}=1-\mathcal{F}$, as a function of the ramp time $t_{\rm ramp}$ and for $\Omega=\Delta_{\rm ph}/4$. Note that the first minimum at $t_{\rm ramp}\approx 0.5\times 2\pi/\Delta_{\rm ph}$ already achieves an error of $\sim 10^{-5}$ (at $T=0$ K). One can follow the procedure for different values of $\Omega$, and choose the value of $t_{\rm ramp}$ according to the minima in $\mathcal{E}$. The result of this optimization is shown in Figs.~\ref{figS4} (b) and (c), where the minimized error $\mathcal{E}_{\rm s-ph}^{\rm min}$ and the ramp time $t_{\rm ramp}$ are plotted as a function of the Rabi frequency $\Omega$. The procedure followed in Sec.~\ref{subsec:ExampleCCDD} is similar, but in that case, both $t_{\rm ramp}$ and $\Omega_\varepsilon$ are optimised to achieve the minimum spin-phonon entanglement error $\mathcal{E}_{\rm s-ph}^{\rm min}$.

\section{Details about the numerical simulations of the CCDD method}\label{app:CCDD}
For the exact numerical simulations of the CCDD method we consider the bare qubit Hamiltonian
\begin{equation} \label{Eq:CDD_exact}
H_{\rm SiV}(t)= \frac{ \Omega(t)}{2} \sigma_x- \Omega_\varepsilon(t)\sin{(\Omega t-\varphi_{\rm r})} \sigma_y,
\end{equation}
which is equivalent to Eq.~\eqref{CCDDHamil} in the main text, except for the additional phase $\varphi_{\rm r}$. This offset is used compensate phase shifts that arise due to the ramps in $\Omega(t)$.

More precisely, we consider the time-dependence for $\Omega(t)$ discussed in Appendix~\ref{app:adiabaticramps} and with a constant $\Omega_\varepsilon(t)=\Omega_\varepsilon$ for $t_{\rm ramp} <t<t_{\rm g}-t_{\rm ramp}$ and $\Omega_\varepsilon(t)=0$ otherwise. In this case the dressed states accumulate an additional phase $\varphi_{\rm r}=\int_0^{t_{\rm ramp}}ds\,\Omega(s)=\Omega t_{\rm ramp}/2$, which must be accounted for when determining the exact shape of the two qubit gate. The resulting gate operation is then given by
\begin{equation}
U_{\rm g}=e^{-i\varphi_x\sigma_x}e^{-i\varphi_z\sigma_z}U^\prime_{\rm g},
\end{equation}
where $U^\prime_{\rm g}= e^{-iH^\prime_{\rm g} t_{\rm g}}$, with $H_{\rm g}^\prime$ given in Eq.~\eqref{eq:Hg_concat} and for a gate time of $t_{\rm g}=\pi/(4J'_{\perp})$. The phases are $\varphi_{x}=\Omega t_{\rm g}/2+\varphi_{\rm r}$ and 
\begin{equation}
\varphi_{z}=\frac{\Omega_\varepsilon }{2}t_{\rm g}  -\frac{\varepsilon}{2}\sin{(\Omega t_{\rm g})}\cos{(\varphi_x/2)}.
\end{equation}

To evaluate the gate error, the ideal target state, e.g. $U_{\rm g}|\Phi\rangle$, is compared (using the state fidelity $\mathcal{F}$)  with the reduced density matrix $\rho_q$ resulting from the application of the master equation in Eq.~\eqref{Eq:Master_Equation}. Note that in this case the Hamiltonian is
\begin{equation}
H(t)=H_{\rm SiV}(t)+ \Delta_{\rm ph}a^\dagger a + \sum_i g_i(\sigma_i^+a +\sigma_i^-a^\dagger),
\end{equation}
with $H_{\rm SiV}(t)$ as defined in Eq.~\eqref{Eq:CDD_exact}.

\section{Mode-continuum limit}\label{app:ContinuumCouplings}
In this Appendix we summarize the derivation of the main analytic results for the coherent and incoherent processes in an extended phononic crystal lattice.  
\subsection{Spin-spin interactions}
In the mode-continuum limit, we can write the sum over all modes $\sum_k$ as the integral $\frac{1}{\Delta k}\int_{k_{\rm min}}^{k_{\rm max}} dk=\frac{N+1}{\pi}\int_0^{\pi} dk$, such that
\begin{equation}
J_{ij}^\parallel = \frac{g_ig_j}{\pi}\int_0^\pi dk  \frac{ \sin{(k i)}\sin{(k j)}}{\Delta_0+4\lambda\sin^2{(k/2)}},
\end{equation}
or,
 \begin{equation}
J_{ij}^\parallel =\frac{g_ig_j}{2\pi\Delta_0(1+x_{\parallel})}\int_0^\pi dk \frac{ \cos{[k(i-j)]}-\cos{[k(i+j)]}}{1-\sin \theta_\parallel\cos{k}},
\end{equation}
where $x_{\parallel}=2\lambda/\Delta_0$ and $\sin \theta_\parallel=x_\parallel/(1+x_\parallel)$. Now, we make use of
\begin{equation}
I(\theta,N)=\int_0^\pi dk \frac{ \cos{(kN)}}{1-\sin{\theta}\cos{k}}=\frac{\pi}{ \cos{\theta}}\, \Big(\frac{2\sin^2{(\theta/2)}}{\sin{\theta}}\Big)^{|N|},
\end{equation} 
to arrive at the result given in Eq.~\eqref{WaveguideCouplings}. Similarly, one obtains the corresponding expression for $J_{ij}^{\perp}$. 

Note that when deriving the results for $J_{ij}^{\parallel,\perp}$, $\sum_m\Lambda_{jm}$ or other coupling terms, we disregard the terms that depend on $|i+j|$, while keeping those depending on $|i-j|$. This is based on the fact that, for $0<\theta< \pi/2$, the following holds,
\begin{equation}
d^{-1}=\frac{2\sin^2{(\theta/2)}}{\sin{\theta}} <1.
\end{equation} 
Since $I(\theta,|i+j|)\propto d^{-|i+j|}$, far from the edges of the crystal, $I(\theta,|i+j|)$ goes to zero. 

\subsection{Deterministic Stark shift}
Similar to the single-mode case, the coupling to many modes will also lead to a deterministic Stark shift of the dressed qubit states given by $\sum_k\Lambda_{jk}(\langle c_k^\dagger c_k\rangle+1/2)$ [see Eq.~(\ref{Eq:SpinSpinHam})]. We assume a sufficiently narrow-band phonon lattice such that $\langle c_k^\dagger c_k\rangle\approx n_{\rm th}$ for all $k$. Then, far from the edges, $|2j|\gg1$, we obtain 
\begin{equation}
\sum_k\Lambda_{jk}\approx \frac{g_i^2 }{4\Delta_0}\sum_{\mu=+,-}\frac{\beta_{\mu}}{\sqrt{1+x_\mu-x_\parallel}\sqrt{1+x_\mu+x_\parallel}} .
\end{equation}
Note that, by taking the single-mode limit, $\lambda\rightarrow 0$ or $x_\parallel\rightarrow0$, we obtain the expression in Eq.~{\eqref{Eq:Starkcoupling}.

\subsection{Stark-shift dephasing}
Following the analysis presented in Appendix~\ref{subapp:StarkDeph}, we can quantify the single-qubit dephasing error produced by Stark shift, in the single-mode case, as 
\begin{eqnarray}
\mathcal{E}^{N=1}_{j}\approx \Lambda_j^2\, {\rm Re}\Big[\int_0^{t_{\rm g}}dt'\int_0^{t'}dt''  \langle \delta \hat{n}_j(t'') \delta \hat{n}_j(0)\rangle\Big].
\end{eqnarray}
We can also apply this expression to the multimode case, with the difference that in that case the local mode at site $j$ couples not only to the bath, but also to the other modes in the system. Consequently, the correlation function $C_{nn}(\tau)=\langle  \delta n_j(\tau)\delta n_j(0)\rangle$ can be expressed as
\begin{equation}
 C_{nn}(\tau)=n_{\rm th}(n_{\rm th}+1)\sum_{k,k'} b^2_{j k}b^2_{jk'} e^{i[ (\Delta_k-\Delta_{k'})-\kappa]\tau},
\end{equation} 
where we assumed all normal modes to be in a stationary thermal state with an average number of phonons $n_{\rm th}$. In the mode-continuum limit, this becomes
\begin{widetext}
\begin{equation}\label{Eq:correlationcontinuum}
C_{nn}(\tau)=\frac{ 4n_{\rm th}(n_{\rm th}+1)}{\pi^2}\int_0^\pi dk\int_0^\pi dk' \sin^2(kj) \sin^2(k'j)e^{i[ (\Delta_k-\Delta_{k'})-\kappa]\tau},
\end{equation} 
Note that, in the small-bandwidth limit $4\lambda\ll 1/t_{\rm g}$, the phase factors $e^{i[ (\Delta_k-\Delta_{k'})\tau}\approx 1 $ for all relevant times, and Eq.~(\ref{Eq:correlationcontinuum}) reduces to Eq.~(\ref{Eq:correlationSinglemode}). On the other hand, in the large-bandwidth limit $4\lambda\gg 1/t_{\rm g}$, where also $\kappa \gg 1/t_{\rm g}$,  we obtain
\begin{equation}\label{Eq:DoubleInt}
\int_0^{t\sim t_{\rm g}} d\tau\, C_{nn}(\tau)\approx \frac{4n_{\rm th}(n_{\rm th}+1)}{\pi^2}\int_0^\pi dk\int_0^\pi dk' \sin^2(ki)\sin^2{(k' i)} \frac{\kappa+i2\lambda(\cos{k}-\cos{k'})}{\kappa^2+4\lambda^2(\cos{k}-\cos{k'})^2}.
\end{equation}
\end{widetext}
For our analysis we only care about the real part of Eq.~(\ref{Eq:DoubleInt}). Also, we can analytically solve
\begin{equation}
\begin{split}
I(k)&=\int_0^\pi dk' \frac{\sin^2{(k' j)}}{\kappa^2+4\lambda^2(\cos{k}-\cos{k'})^2}\\
&=\frac{\pi}{4\lambda\kappa} {\rm Im}\big[\tan(z) \{1- (\alpha_k)^{|2j|}\} \big],
\end{split}
\end{equation}
where $\alpha_k=2\sin^2 (z/2)/\sin(z)$  and $\sin(z)=2\lambda/(2\lambda\cos{k}-i\kappa)$.
Assuming that site $j$ is far from the edges, we have $I(k)\approx \frac{\pi}{4\lambda\kappa}[\tan(z) -c.c.]/(2i)$. Under the same approximation we find that
\begin{widetext}
\begin{equation}
\begin{split}
{\rm Re}\Big[\int_0^{t\sim t_{\rm g}} d\tau \,C_{nn}(\tau)\Big]=\frac{n_{\rm th}(n_{\rm th}+1)}{\pi \lambda} \int_0^\pi dk\sin^2(kj)\, I(k)\,\Big(\frac{\pi}{4\lambda\kappa}\Big)^{-1}\approx\frac{n_{\rm th}(n_{\rm th}+1)}{2\pi \lambda}\int_0^\pi dk\, I(k)\,\Big(\frac{\pi}{4\lambda\kappa}\Big)^{-1}.
\end{split}
\end{equation}
\end{widetext}
In summary, the Stark-dephasing error in the multimode case is
\begin{equation}
\begin{split}\label{Eq:Realpart}
\mathcal{E}^{N\rightarrow\infty}_{j}&\approx \Lambda_j^2 t_{\rm g,m}\, {\rm Re}\left[\int_0^\tau d\tau\langle  \delta n_j(\tau)\delta n_j(0)\rangle\right]\\
&\approx   \Gamma_{ii}^zt_{\rm g,m}\, {\rm Im}\left[ \int_0^{\pi} \frac{dk}{\sqrt{(\cos k-i \kappa/2\lambda)^2-1}}\right],
\end{split}
\end{equation}
where $t_{\rm g,m}$ is the gate time in the multimode case, and $\Gamma_{i,j}^z=\Lambda_{i}\Lambda_{j}n_{\rm th}(n_{\rm th}+1)/(2 \pi \lambda)$ is a characteristic dephasing rate. While for two qubits coupled via an on-site mode the gate time is $t_{\rm g}\approx J_{i,i}^{-1}$, for two qubits sitting in neighboring sites the gate time will be $t_{\rm g,m}\approx J_{i,i+1}^{-1}\sim 10 J_{i,i}^{-1}$.

We numerically solve the integral in Eq.~(\ref{Eq:Realpart}), $\mathcal{I}$, for the parameters considered in the main text, i.e., $\omega_{\rm ph}=0.05 \Delta_0$, and $Q=10^6$, and for a bandwidth $2\lambda/\Delta_0\approx1$ ($\kappa/2\lambda\approx \kappa/\Delta_0 \approx 2\times10^{-5}$) and obtain $\rm Im[\mathcal{I}]\approx 10$. The corresponding single-qubit dephasing error is then
\begin{equation}
\mathcal{E}^{N\rightarrow\infty}_{j}\approx \frac{\Lambda_i^2}{\pi J_{i,i}\lambda} n_{\rm th}(n_{\rm th}+1)\times 100,
\end{equation}
while in the single-mode case we found the scaling  
\begin{equation}
\mathcal{E}^{N=1}_{j}\approx \frac{\Lambda_i^2}{ J_{i,i}\kappa} n_{\rm th}(n_{\rm th}+1)
\end{equation}
instead. Therefore, in the waveguide limit the thermal dehasing error is reduced by a factor of about $(\kappa/\pi\lambda)\times 100\approx 10^{-3}$.

-----------------------------------------------------------------------

\end{document}